\documentclass[prd,preprint,superscriptaddress,preprintnumbers,eqsecnum,showpacs,nofootinbib,nobibnotes]{revtex4}
\usepackage{amsfonts,bm}
\usepackage{amsmath}
\usepackage{amssymb}
\usepackage{graphicx}
\usepackage{subfigure}

\newcommand{\be}{\begin{equation}}
\newcommand{\bea}{\begin{eqnarray}}
\newcommand{\ee}{\end{equation}}
\newcommand{\eea}{\end{eqnarray}}
\newcommand{\bpi}{\begin{picture}}
\newcommand{\bce}{\begin{center}}
\newcommand{\epi}{\end{picture}}
\newcommand{\ece}{\end{center}}

\def\s#1{{\scriptscriptstyle #1}}

\def\gtreeb{\widetilde{\Gamma}^{(0)}}
\def\gfullb{\widetilde{\Gamma}}

\def\diff{{\rm d}}

\begin{document}

\title{Gluon mass through ghost synergy}

\author{A.~C. Aguilar}
\affiliation{Federal University of ABC, CCNH, 
Rua Santa Ad\'elia 166,  CEP 09210-170, Santo Andr\'e, Brazil}

\author{D. Binosi}
\affiliation{European Centre for Theoretical Studies in Nuclear
Physics and Related Areas (ECT*) and Fondazione Bruno Kessler, \\Villa Tambosi, Strada delle
Tabarelle 286, 
I-38123 Villazzano (TN)  Italy}

\author{J. Papavassiliou}
\affiliation{\mbox{Department of Theoretical Physics and IFIC,  
University of Valencia and CSIC},
E-46100, Valencia, Spain}

\begin{abstract}

In this work we 
compute, at the ``one-loop-dressed'' level,  
the nonperturbative contribution of the ghost loops 
to the self-energy of the gluon propagator,   
in the Landau gauge. 
This is accomplished 
within the  PT-BFM formalism, which guarantees the gauge-invariance of the 
emerging answer. In particular, the contribution of the ghost-loops is automatically transverse, by virtue of the 
QED-like Ward identities satisfied in this framework.  
Using  as nonperturbative input 
the  available  lattice data  
for  the ghost dressing function, 
we  show   that  the   ghost  contributions 
have a rather sizable effect on the overall shape of the gluon propagator,
both for $d=3,4$. 
Then, by  exploiting  a recently  introduced  dynamical equation  
for the effective gluon mass, whose solutions depend crucially on the 
characteristics of the gluon propagator at intermediate energies,  
we show that  if the ghost loops are 
removed from the gluon propagator then 
the gluon mass vanishes. These findings strongly suggest that, 
at least at the level of the Schwinger-Dyson equations,  
the effects of gluons and ghosts are inextricably connected, 
and must  be combined suitably in order to  
reproduce the results
obtained  in the recent lattice  simulations.

\end{abstract}

\pacs{
12.38.Aw,  
12.38.Lg, 
14.70.Dj 
}

\maketitle

\section{Introduction}
 
Our understanding of  the infrared (IR) properties of  
the fundamental Green's functions of Yang-Mills theories 
has improved considerably in the last few years, 
due to a variety of parallel efforts in   
lattice simulations~\cite{Cucchieri:2007md,Cucchieri:2007rg,Cucchieri:2009zt,
Cucchieri:2011ga,Cucchieri:2011um,Cucchieri:2003di,Cucchieri:2010xr,
Bowman:2007du,Bogolubsky:2007ud,Bogolubsky:2009dc,Oliveira:2008uf,Oliveira:2009eh},
Schwinger-Dyson 
equations (SDEs)~\cite{Aguilar:2008xm, Aguilar:2010zx, Binosi:2009qm, RodriguezQuintero:2011vw,
RodriguezQuintero:2010wy, Boucaud:2010gr, Boucaud:2008gn, Boucaud:2008ji}, 
functional methods~\cite{Braun:2007bx,Szczepaniak:2010fe},
and algebraic  techniques~\cite{Dudal:2008sp,Dudal:2010tf,Dudal:2011gd,Kondo:2011ab}.
The majority of the aforementioned studies have focused on the low-momentum 
behavior the   gluon   and   ghost
propagators, which 
can be directly or indirectly 
related to some of the most fundamental nonperturbative phenomena of QCD,
such as quark confinement, dynamical mass generation, and chiral symmetry breaking.

It is by now well-established that, in the  Landau gauge, the
lattice yields a gluon propagator  and a ghost dressing function that are
finite  in the  IR (in $d=3,4$)~\cite{Aguilar:2008xm,Aguilar:2010zx,Boucaud:2008ji}. 
 Evidently, these  lattice results furnish 
strong support to the idea of dynamical  gluon mass generation  
through the well-known Schwinger mechanism~\cite{Jackiw:1973tr,Cornwall:1973ts,Eichten:1974et}, 
as proposed by Cornwall and others~\cite{Cornwall:1981zr, Bernard:1981pg}. 
On the other hand, these important lattice findings have 
motivated the critical revision of the original Gribov-Zwanziger confinement scenario, 
leading to the formulation of its ``refined'' version~\cite{Dudal:2008sp}. 
In addition,  
the ``ghost-dominance'' picture of QCD~\cite{Alkofer:2000wg,Fischer:2006ub}, whose 
theoretical cornerstone  
has been the existence of a divergent (``IR-enhanced'') ghost dressing function, 
is at odds with the above lattice results, and, at least in this strict formulation,   
has been  practically ruled out (in the Landau gauge, and for $d=3,4$)~\cite{Aguilar:2008xm,Aguilar:2010zx,Boucaud:2008ji}.

This last statement, 
however, does not necessarily mean that 
the ghost has been relegated to a marginal role in the  QCD dynamics. 
In fact, compelling evidence to the contrary 
has emerged from  
detailed studies of the gap equation that controls
the breaking of chiral symmetry and the dynamical generation 
of a constituent quark mass~\cite{Fischer:2003rp,Aguilar:2010cn}. Specifically, the 
proper inclusion of   the corresponding ghost sector 
(essentially the ghost dressing function and the quark-ghost kernel) 
is crucial for 
obtaining a realistic symmetry breaking pattern, with quark masses
in the phenomenologically relevant range. 
The main lesson that can be drawn from the above studies is that 
even a finite ({\it i.e.}, ``non-enhanced'') ghost sector may 
have a strong numerical impact,  
at least in the framework of the SDEs, and affect nontrivially 
the realization of various underlying dynamical mechanisms~\cite{Aguilar:2010cn}.
In fact, 
for the concrete case of the quark gap equation, the
ghost contributions provide the necessary enhancement to the 
kernel of the gap equation precisely in the 
range of momenta  around \mbox{1 GeV}, 
which is the most relevant for obtaining the right type 
of quark mass solutions~\cite{Roberts:1994dr}.

Given the importance of the ghost sector for the dynamical generation 
of a constituent quark mass, 
it is natural to ask whether a similar situation applies  
in the case of the dynamical generation of an effective gluon mass. 
The main purpose of 
the present article is to address in detail this important question.

\begin{figure}[!t]
\includegraphics[scale=1.275]{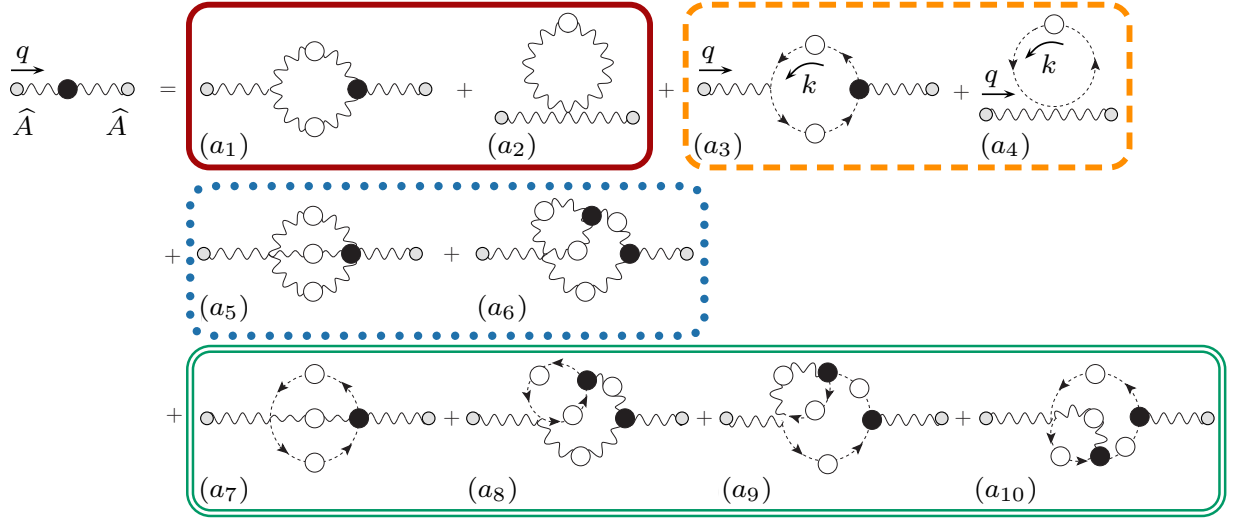}
\caption{\label{fig1}The SDE corresponding to the PT-BFM    gluon     self-energy
$\Pi^{ab}_{\mu\nu}(q)$. The graphs inside each box 
form a gauge invariant subgroup,
furnishing an individually transverse contribution. White
(black) circles denote full propagators (vertices).}
\end{figure}

This problem is technically rather subtle, and hinges on the 
ability to treat self-consistently  various field theoretic ingredients. 
To that end, we will employ the general  formalism based  on the  pinch technique
(PT)~\cite{Cornwall:1981zr,Cornwall:1989gv,Binosi:2002ft,Binosi:2003rr,Binosi:2009qm}  
and  the  background  field
method  (BFM)~\cite{Abbott:1980hw},
which is particularly suited for dealing 
precisely with this type of problem. Specifically, the truncation scheme 
based on the PT-BFM formalism~\cite{Aguilar:2006gr,Binosi:2007pi,Binosi:2008qk} allows 
for subtraction of the ghost contributions 
to the gluon self-energy in a  physically meaningful way 
({\it i.e.}, without introducing gauge artifacts). 
Indeed, in the conventional SDE formulation, any attempt to 
isolate the ghost contributions is bound to 
interfere with the transversality of the resulting gluon 
self-energy; this can be seen already at the one-loop level,
where only the sum of the gluon and ghost diagrams (but not their individual 
contributions) is   
transverse. Instead, as was first pointed out in the classic 
paper by Abbott~\cite{Abbott:1980hw}, 
the calculation of the same diagrams  
using the BFM Feynman rules 
gives rise to two transverse contributions. This crucial 
property persists unaltered at the level of the SDE for the gluon self-energy:  
the SDE is composed by concrete subsets 
of ``one-''  and ``two-loop dressed'' diagrams,  which are separately
transverse, {\it e.g.}, $q_{\mu}\Pi^{\mu\nu}_c(q)=0$, 
where $\Pi^{\mu\nu}_c(q) = (a_3)^{\mu\nu} +( a_4)^{\mu\nu}$ - see  Fig.~\ref{fig1}.
Therefore, one  can study the  individual contribution of
the different  blocks [in this case $\Pi^{\mu\nu}_c(q)$]
to the full gluon self-energy,  
without compromising the transversality of   the  answer
(these points have been addressed in great detail 
in~\cite{Aguilar:2006gr,Binosi:2009qm,Binosi:2007pi,Binosi:2008qk}) 

Given that within the PT-BFM framework the ghost contributions to the gluon
self-energy 
may be disentangled gauge invariantly,
the next step will be to compute this particular  
contribution  nonperturbatively, 
and then subtract it out 
from the full gluon propagator obtained from the lattice.
The basic operating assumption underlying this analysis is that 
the gluon propagator found on the lattice {\it coincides} with that 
obtained from the solution of the full SDE series. 
Then, instead of solving the SDE series without the loops contained in  $\Pi^{\mu\nu}_c(q)$ 
to determine the resulting gluon propagator (technically an impossible task at the moment), 
we compute nonperturbatively only the contribution of $\Pi^{\mu\nu}_c(q)$ 
and subtract it from the gluon propagator obtained from the lattice.
The nonperturbative computation of the aforementioned ghost contribution  $\Pi^{\mu\nu}_c(q)$
[graphs $(a_3)$ and $(a_4)$ in Fig.~\ref{fig1}]
proceeds through the following main steps. 

\begin{enumerate}

\item[\it{(i)}] We introduce a suitable  Ansatz for the 
full (background) gluon-ghost vertex, $\gfullb_\mu$, [the black circle in graph $(a_3)$] 
which satisfies automatically ({\it i.e.}, by construction)  
the all-order Ward identity given in Eq.~(\ref{WI}).  
This is an indispensable requirement  
for maintaining the gauge invariance (transversality) of the answer. 
The $\gfullb_\mu$ obtained from this procedure 
[given in Eq.(\ref{va})] 
is expressed entirely in terms of the ghost propagator. 
As a result, the only quantity 
appearing finally inside the graphs $(a_3)$ and $(a_4)$ is 
the ghost propagator (or its dressing function).

\item[\it{(ii)}]  We invoke the so-called ``seagull identity'', 
given in Eq.~(\ref{seagull}), 
which enforces 
the cancellation of all sorts of seagull-type contributions, leading to the 
absence of quadratic divergences~\cite{Aguilar:2009ke}.  
Specifically, by means of this identity 
the purely seagull contribution of graph $(a_4)$ cancels in its entirety 
against a term obtained from $(a_3)$. 

\item[\it{(iii)}]  The remaining expression for $\Pi^{\mu\nu}_c(q)$ 
is renormalized {\it subtractively}, according to the 
rules of the momentum-subtraction (MOM) prescription.

\item[\it{(iv)}]  The renormalized expression for $\Pi^{\mu\nu}_c(q)$ is then 
computed numerically, by substituting for the (infrared finite) 
ghost dressing function, appearing 
inside the integrals, the available lattice data for 
this quantity~\cite{Cucchieri:2007md,Cucchieri:2011um,Bogolubsky:2007ud}.  

\item[\it{(v)}] 
The latter contribution is subtracted 
from the entire gluon propagator obtained from the lattice, according to the 
formula given in Eq.~(\ref{me}).

\end{enumerate}

The results 
turn out to be  rather striking (see the right panels of  
Figs.~\ref{GdressandChi-4dSU3}, \ref{RTandGluon-4dSU2} and \ref{RTandGluon-3d}): 
the gluon propagator without $\Pi^{\mu\nu}_c(q)$ 
is significantly different from the full one. 
In addition, the results suggest a strong dependence on the
space-time dimensionality: the effect of removing ghosts becomes 
considerably more enhanced as the space-time dimensionality is lowered.

At  this  point one  can turn to the main question of this work, 
and study  what would  happen to the gluon mass  
if the ghost contributions, computed in the previous steps, 
were to be removed from the full gluon propagator obtained from the lattice. 
This question  can be  addressed in quantitative detail by means 
of the integral  equation, derived recently in~\cite{Aguilar:2011ux}, 
which describes the  evolution ({\it i.e.}, momentum-dependence) of  the
dynamical gluon  mass,~$m^2(q^2)$. 
This particular equation, given in Eq.~(\ref{me-final}), 
contains as its main ingredient the full gluon propagator~ 
$\Delta$, which practically determines the form 
of its kernel.
The detailed analysis of an approximate version of 
Eq.~(\ref{me-final}) carried out in~\cite{Aguilar:2011ux} reveals 
that the existence of physically acceptable 
solutions hinges crucially on the shape of the gluon propagator
in the entire range of physical (Euclidean) momenta, and in particular 
on the precise behavior that $\Delta$ 
displays in the region between \mbox{(1-5) $\rm GeV^2$}. 
Specifically, in order for the gluon mass to be positive definite,
 the first derivative 
of the quantity $ q^2 \Delta(q^2)$ (the ``gluon dressing function'')
must furnish a sufficiently {\it negative} 
contribution in the aforementioned range of momenta. 
Note that, as was shown in~\cite{Aguilar:2011ux},  
the full $\Delta$ obtained from the lattice has indeed this particular property, 
giving rise [when inserted into  Eq.~(\ref{me-final})] 
to a dynamically generated gluon mass with the expected characteristics. 
Evidently, the main effect of removing the ghost contributions contained in  $\Pi^{\mu\nu}_c(q)$ 
from $\Delta(q^2)$ is to restrict significantly the 
negative area displayed by the (ghostless) $\Delta(q^2)$ (see the right panels of  Fig.~\ref{Kernel-all}), 
a fact which, in turn, leads to the 
vanishing of the gluon mass, {\it i.e.}, the homogeneous Eq.~(\ref{me-final}) 
can only admit the trivial solution $m^2(q^2)=0$. 

Interestingly enough, 
this  result  appears to be completely
analogous to what happens in  the case of chiral symmetry breaking, 
where failure to include ghost contributions into  
the  gap  equation [the quark analogue of Eq.~(\ref{me-final})] 
prevents the dynamical generation of 
a constituent quark mass~\cite{Aguilar:2010cn}.
Thus, in the picture of QCD emerging from this analysis,  
ghosts  and  gluons must be in a state of harmonious synergy  
in order for a mass gap to be produced, 
regardless of the nature of 
the fundamental particle in question (gluon or quark).

The article is organized as follows. In Section~\ref{disen} we study the
general properties  of the ghost sector  in the Landau  gauge and explain 
how  it is  possible within  the PT-BFM  framework to  disentangle 
gauge-invariantly the
(one-loop dressed) ghost contributions,  $\Pi^{\mu\nu}_c(q)$, from the full gluon propagator. 
In Section~\ref{nf}
we derive the non-perturbative expression that determines $\Pi^{\mu\nu}_c(q)$
solely in terms of the ghost propagator and the coupling constant.
In Section~\ref{num} we evaluate numerically the expressions for $\Pi^{\mu\nu}_c(q)$ 
derived in the previous section, 
using as input the ghost dressing function 
obtained in recent lattice 
simulations~\cite{Cucchieri:2007md,Cucchieri:2011um,Bogolubsky:2007ud}. Next,  
we determine how the removal of $\Pi^{\mu\nu}_c(q)$ 
affects the  overall   shape  of   the   resulting   gluon propagator,
for three different cases: \mbox{$d=4$} and \mbox{$N=2,3$}, as well as \mbox{$d=3$} and \mbox{$N=2$}, 
where $d$ is the dimensionality of space-time  and $N$ is 
the number of  colors [corresponding to the gauge group $SU(N)$]. 
In Section~\ref{DMG} we  turn to the main question of the present 
work, and study in detail how  the  kernel  of  the  dynamical  integral  equation
governing  the   gluon  mass  gets  modified  after removing 
the aforementioned ghost contributions. 
Finally, our  conclusions  are presented in Section~\ref{concl}.

\section{\label{disen} Gauge invariant subtraction of ghost loops}
In this section we first 
derive the formula that will determine the residual 
gluon propagator obtained from the full gluon propagator after 
removing from the latter the 
``one-loop dressed'' ghost contributions, given by diagrams $(a_3)$ and $(a_4)$ in Fig.~\ref{fig1}. 
Then, we work out the nonperturbative expression that determines the 
aforementioned ghost contribution in terms of integral involving the 
ghost dressing function.

The first important fact to recognize is 
the transversality of the ghost contributions to be removed.
Specifically, denoting their sum by 
\be
\Pi_c^{\mu\nu}(q)=(a_3)^{\mu\nu}+(a_4)^{\mu\nu} \,,
\label{Pica}
\ee
we have that 
\bea
(a_3)_{\mu\nu}&=& -g^2C_A
\int_k\!\gtreeb_{\mu}(k,q,-k-q)D(k)D(k+q)\gfullb_{\nu}(k+q,-q,-k), 
\nonumber \\
(a_4)_{\mu\nu}&=&2 g^2C_A g_{\mu\nu}\int_k\!D(k).
\label{gh-1ldr}
\eea
In the equations above, $D^{ab}(q^2)=\delta^{ab}D(q^2)$ denotes the full ghost propagator, 
defined in terms of the ghost dressing function $F$ as
\be
D(q^2)= \frac{F(q^2)}{q^2},
\label{ghdr}
\ee
while $\gfullb_{\mu}$ represents the three-particle vertex describing 
the interaction of the background gluon with a ghost and an antighost, 
with (all momenta entering)
\be
i\Gamma_{c^b \widehat{A}^a_\mu \bar c^c}(r,q,p)=gf^{acb}\widetilde{\Gamma}_{\mu}(r,q,p); \qquad \widetilde{\Gamma}^{(0)}_{\mu}(r,q,p)=(r-p)_\mu.
\ee
Finally, $C_A$ is the Casimir eigenvalue of the adjoint representation	
[$C_A=N$ for $SU(N)$], and we have introduced the $d$-dimensional 
integral measure (in dimensional regularization) according to
\be
\int_{k}\equiv\frac{\mu^{\epsilon}}{(2\pi)^{d}}\!\int\!\mathrm{d}^d k,
\label{dqd}
\ee
with $\mu$ the 't Hooft mass, and $\epsilon=4-d$.

Then, by virtue of the PT-BFM Ward identity 
\be
iq^\mu\gfullb_\mu(r,q,p)=D^{-1}(r)-D^{-1}(p),
\label{WI}
\ee
it is immediate to establish the transversality of $\Pi_c^{\mu\nu}(q)$, namely~\cite{Aguilar:2006gr}
\be
q_{\mu}\Pi_c^{\mu\nu}(q) =0.
\ee
Let us now denote by $\Pi^{\mu\nu}_r(q)$ the sum 
of the remaining subsets of diagrams in  
Fig.~\ref{fig1}, {\it i.e.}, both the gluon one- and two-loop dressed diagrams, 
as well as two-loop dressed ghost diagrams, 
\be
\Pi^{\mu\nu}_r(q) = \sum_{\substack{i=1\\i\neq3,4}}^{10} (a_i)^{\mu\nu}\,.
\ee

Again, due to the special Ward identities satisfied by the PT-BFM vertices, 
$\Pi^{\mu\nu}_r(q)$ is also transverse, and, of course, so is the full self-energy $\Pi^{\mu\nu}(q)$, given simply by 
\be
\Pi^{\mu\nu}(q) = \Pi^{\mu\nu}_r(q) + \Pi_c^{\mu\nu}(q) .
\ee

The SDE for the full gluon propagator in the Landau gauge of the PT-BFM scheme assumes then the form 
\be
\Delta^{-1}(q^2)P^{\mu\nu}(q)=\frac{q^2P^{\mu\nu}(q)+i \left[\Pi^{\mu\nu}_r(q)+\Pi^{\mu\nu}_c(q)\right]}{\left[1+G(q^2)\right]^2},
\label{gSDE}
\ee 
where the gluon propagator $\Delta_{\mu\nu}(q)$ is defined as (we suppress color indices) 
\be
\Delta_{\mu\nu}(q)=-i\Delta(q^2)P_{\mu\nu}(q); \qquad
P_{\mu\nu}(q)=g_{\mu\nu}-\frac{q_\mu q_\nu}{q^2},
\ee
The function $G$ appearing in (\ref{gSDE}) is the form factor associated with $g_{\mu\nu}$
in the Lorentz decomposition of the auxiliary two-point function $\Lambda$, given by 
\bea
\Lambda_{\mu\nu}(q)&=&-ig^2C_A\int_k\!\Delta_\mu^\sigma(k)D(q-k)H_{\nu\sigma}(-q,q-k,k)\nonumber\\
&=&g_{\mu\nu}G(q^2)+\frac{q_\mu q_\nu}{q^2}L(q^2).
\eea

\begin{figure}[!t]
\includegraphics[scale=.75]{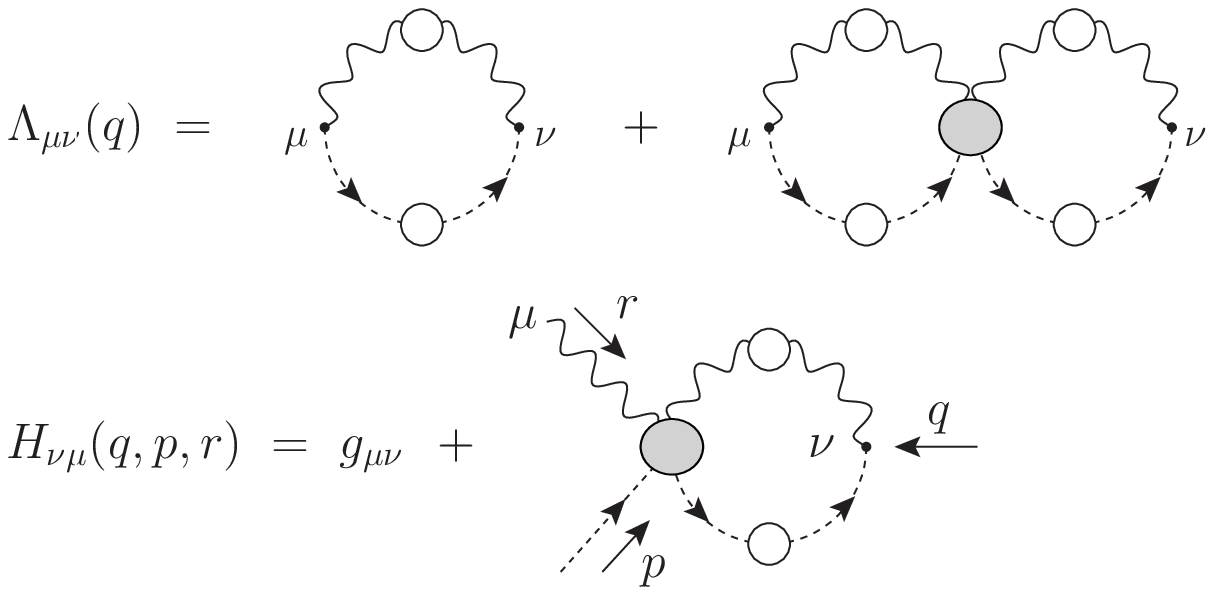}
\caption{\label{fig3}Definitions and conventions of the auxiliary 
functions $\Lambda$ and $H$. Gray blobs denote  1-PI  kernels (with respect to vertical cuts).}
\end{figure}

This latter function, together with the auxiliary function $H$, are diagrammatically represented  in Fig.~\ref{fig3}; 
also notice that  $H$ is  related to the (conventional) gluon-ghost vertex by the identity
\be
p^\nu H_{\nu\mu}(p,r,q)+\Gamma_{\mu}(r,q,p)=0,
\ee
and that, in the (background) Landau gauge,  
 the following all order relation holds~\cite{Grassi:2004yq,Aguilar:2009pp}
\be
F^{-1}(q^2)=1+G(q^2)+L(q^2).
\label{funrel}
\ee 

Now, let us return to Eq.~(\ref{gSDE}), 
and define in a  completely analogous way the quantity $\Delta_r(q^2)$, given by 
\be
\Delta_r^{-1}(q^2)P^{\mu\nu}(q)=\frac{q^2P^{\mu\nu}(q)+i\Pi^{\mu\nu}_r(q)}{\left[1+G(q^2)\right]^2}. 
\label{Dr}
\ee
Evidently, $\Delta_r$ represents the propagator 
obtained by subtracting out (gauge invariantly) from the full propagator 
$\Delta$ the one-loop dressed ghost contributions. 
Then, taking the trace of both Eqs.~(\ref{gSDE}) and (\ref{Dr}), 
defining the trace of $\Pi^{\mu\nu}_c(q)$ as 
\be
\Pi_c(q^2) \equiv \Pi^{\mu}_{c\,\mu}(q) , 
\label{Pic}
\ee
 and solving for $\Delta_r$, we arrive at 
\be
\Delta_r(q^2)=\Delta(q^2)\left\{1-\frac{i\Delta(q^2)\Pi_c(q^2)}{(d-1)\left[1+G(q^2)\right]^2}\right\}^{-1},
\label{me}
\ee
which represents our master formula.

In order to obtain the behavior of the  
propagator $\Delta_r(q^2)$ from Eq.~(\ref{me}) we will 
{\it (i)} identify the full gluon propagator $\Delta(q^2)$
with that obtained from the lattice, and {\it (ii)} 
determine nonperturbatively the quantity $\Pi_c$ 
from Eqs.~(\ref{gh-1ldr}) and (\ref{Pic}), and evaluate it numerically 
using as input the lattice results 
for the ghost dressing function $F(q^2)$. 
These points will be the subject of the next two sections. 

\section{\label{nf}The nonperturbative expression  for $\Pi_c(q^2)$.}

To accomplish step {\it (ii)} above, we first need to introduce an 
Ansatz for the fully-dressed ghost vertex $\gfullb_{\mu}$, appearing 
in graph $a_3$ of Eq.~(\ref{gh-1ldr}), which satisfies 
the crucial Ward identity of Eq.~(\ref{WI}) (this general procedure is known 
as the ``gauge-technique~\cite{Salam:1963sa}). 
The required Ansatz is easily constructed  from that derived in~\cite{Ball:1980ay} 
for the case of scalar QED case, requiring the absence of kinematic or dynamical singularities. 
It reads 
\be
\gfullb_\mu(r,q,p)=i\frac{(r-p)_\mu}{r^2-p^2}\left[D^{-1}(p^2)-D^{-1}(r^2)\right],
\label{va}
\ee
and evidently satisfies Eq.~(\ref{WI}) when contracted with $q^{\mu}$. 
Obviously, the procedure of reconstructing the vertex by ``solving'' its Ward identity (known in general as ``gauge technique'') 
leaves the transverse (automatically conserved) part of the vertex undetermined~\cite{Salam:1963sa, Kizilersu:2009kg}. In 
this case this term has the form ${\cal A}(r,q)\left[(r\cdot q)p_\mu-(p\cdot q)r_\mu\right]$. 
This particular term vanishes as $q \to 0$, provided that 
the form factor ${\cal A}(r,q)$ does not diverge too strongly in that limit, 
which we will assume in what follows.
Under this assumption, the transverse part of the vertex is subleading in the IR. On the other hand, its omission is known 
to affect the renormalization properties 
of the resulting SDE, a fact that forces one to renormalize subtractively 
instead of multiplicatively (see below).

Substituting~(\ref{va}) in the first equation of~(\ref{gh-1ldr}) and taking the trace,
it is relatively straightforward to obtain the result
\be
\Pi_c(q^2) = g^2C_A \left[4T(q) - q^2R(q)\right],
\label{PicTR}
\ee
where 
\bea
R(q)&=&\int_k\!\frac{D(k+q)-D(k)}{(k+q)^2-k^2},\nonumber \\
T(q)&=&\int_k\!k^2\frac{D(k+q)-D(k)}{(k+q)^2-k^2}+\frac d2\int_k\!D(k).
\label{RandT}
\eea
To further evaluate $\Pi_c(q^2)$, we must invoke 
the so-called ``seagull-identity''~\cite{Aguilar:2009ke},  
\be
\int_k\! k^2\frac{\partial{f}(k^2)}{\partial k^2}+\frac d2\int_k\!f(k^2)=0, 
\label{seagull}
\ee 
valid in dimensional regularization, which enforces the cancellations of all seagull-type of divergences.
This identity guarantees the (nonperturbative) masslessness of the photon in scalar QED 
[by setting $f(k^2) \to {\mathcal D}(k^2)$, where ${\mathcal D}(k^2)$ is the full massive scalar propagator], as well as  
the absence of quadratic divergences from the SDE determining the dynamical 
gluon mass  [by equivalently setting $f(k^2) \to \Delta(k^2)$]~\cite{Aguilar:2009ke}.  

For the case at hand, what we want to guarantee is that $\Pi_c(0) =0$; this must be indeed so, because 
the ghost-loop giving rise to $\Pi_c(q^2)$ has no {\it direct} knowledge of the 
mass generating mechanism, namely the fact that $\Delta^{-1}(0) = m^2(0)$. 
The easiest way to appreciate this is by recalling that the mechanism responsible for 
endowing the gluon with a dynamical mass relies on the presence of massless poles in the 
nonperturbative tree-gluon
[the black circle in graph $(a_1)$ of Fig.~\ref{fig1}], whereas the ghost vertex has the usual structure
[note the absence of poles in the Ansatz of Eq.~(\ref{va})]~\cite{Aguilar:2011ux}. 

Evidently, in the limit  $q\to0$, the term $q^2R(q)$ vanishes, and so does $T(q)$, since 
\bea
T(q)\ \stackrel{q\to0}{\to}\ T(0) &=&
\int_k\!k^2\frac{\partial D(k^2)}{\partial k^2}+\frac d2\int_k\!D(k),
\nonumber \\
&=& 0 , 
\label{T0}
\eea
where in the last step we have employed Eq.~(\ref{seagull}), with \mbox{$f(k^2) \to D(k^2)$}.

In addition, note that the  perturbative (one-loop) version of 
the terms $R(q)$ and $T(q)$, obtained from Eq.~(\ref{RandT}) by setting  
\mbox{$D(k^2) = 1/k^2$}, is given by 
\bea 
R^{(1)}(q)&=& -\int_k \,\frac{1}{k^2 (k+q)^2},\nonumber \\
T^{(1)}(q)&=& ( d/2 -1) \int_k\,\frac{1}{k^2} .
\label{RandTpert}
\eea
Evidently, due the dimensional regularization result \mbox{$\int_k\! k^{-2}=0$} , we have that  \mbox{$T^{(1)}(q)=0$}, 
and in the limit \mbox{$q\to0$}, \mbox{$q^2R^{(1)}(q)$} vanishes (in \mbox{$d=3,4$}).

It is clear that, when $d=4$,  $R(q)$ is 
ultraviolet divergent, and must be properly renormalized, by introducing in the original Lagrangian the  
appropriate counterterm or wave-function renormalization (the need to renormalize is seen explicitly 
already at the level of $R^{(1)}(q)$, which diverges logarithmically). 
The (nonperturbative) renormalization of $\Pi_c(q^2)$ that we will employ proceeds as follows. 
First of all, as happens almost exclusively 
at the level of SDEs, the renormalization must be carried out subtractively instead of 
multiplicatively. The main reason for that is the mishandling of overlapping divergences 
due to the ambiguity inherent in the gauge-technique construction of the vertex, related with the 
unspecified transverse part~\cite{Curtis:1990zs}. 

The (subtractive) renormalization must be carried out at the level of (\ref{gSDE}).
Specifically (setting directly $d=4$),  
\be
\Delta^{-1}(q^2)=\frac{Z_A q^2+\frac{i}{3}\left[\Pi_r(q) +\Pi_c(q)\right]}{\left[1+G(q^2)\right]^2},
\label{rgSDE}
\ee 
where
the renormalization constant $Z_A$ is fixed in the MOM scheme through the
condition $\Delta^{-1}(\mu^2) = \mu^2$. 
This condition, when applied at the level of Eq.~(\ref{rgSDE}), allows one to express $Z_A$ as
\be
Z_A=[1+G(\mu^2)]^2 - \frac{i}{3\mu^2} \left[\Pi_{g}(\mu) + \Pi_{c}(\mu)\right].
\label{z1}
\ee
Now, as is well-known~\cite{Aguilar:2009nf,Aguilar:2009pp},  
the validity of the BRST-driven relation (\ref{funrel}) before and after renormalization prevents  
$G(\mu^2)$ from vanishing when,  according to the MOM prescription, $F(\mu^2) =1$; instead, we must impose that $G(\mu^2) =- L(\mu^2)$. 
However, given that  $L(x)$ is considerably smaller 
than $G(x)$ in the entire range of momenta, we can use the approximation $1+G(\mu^2) \approx F^{-1}(\mu^2)=1$,   
without introducing an appreciable numerical error.  Thus,  
we obtain the following approximate equation for $Z_A$ 
\be
Z_A= 1 - \frac{i}{3\mu^2}\left[\Pi_{r}(\mu) + \Pi_{c}(\mu) \right].
\label{z12}
\ee
Finally, substituting Eq.~(\ref{z12}) into Eq.~(\ref{rgSDE}), and defining (in a natural way)   
the renormalized $\Delta^{-1}_r(q^2)$ as 
\bea
\Delta^{-1}_r(q^2)= \frac{ q^2 + \frac{i}{3}\left[\Pi_{r}(q) - (q^2/\mu^2)\Pi_{r}(\mu)\right]}{[1+G(q^2)]^2},
\label{deltag_re}
\eea 
the renormalized version of the master formula (\ref{me}) will read
\be
\Delta_{r}^{-1}(q^2) = \Delta^{-1}(q^2) - 
\frac{i}{3}\frac{\left[\Pi_{c}(q) - (q^2/\mu^2)\Pi_{c}(\mu)\right]}{[1+G(q^2)]^2}.  
\label{rme}
\ee
Evidently (\ref{rme}) is obtained from (\ref{me}) by replacing  $\Delta^{-1}(q^2) \to \Delta^{-1}_{R}(q^2)$  
(``$R$'' for ``renormalized''), and  $\Pi_{c}(q) \to  \Pi_{c, R}(q)$, where  
\be
\Pi_{c, R}(q) = \Pi_{c}(q) - (q^2/\mu^2)\Pi_{c}(\mu).
\label{rpic}
\ee

As an elementary check, note that the application of the last formula at one loop yields  
\bea  
\Pi_{c, R}^{(1)}(q) &=& -g^2C_A q^2 [R^{(1)}(q)- R^{(1)}(\mu)]
\nonumber \\
&=& \frac{i g^2C_A}{16\pi^2} q^2 \ln\left(q^2/\mu^2\right), 
\eea
which is the standard one-loop result of the PT-BFM~\cite{Cornwall:1989gv,Abbott:1980hw}, renormalized in the MOM scheme. 

For the ensuing numerical treatment of $R(q)$ and $T(q)$
carried out in the next section, it is advantageous to have the 
crucial property $T(0)=0$ a priori built in, 
in order to avoid possible deviations due to minor numerical instabilities. 
To that end, we introduce the quantity $\overline{T}$
\bea
\overline{T}(q)&=&T(q)-T(0)\nonumber \\
&=&\int_k\!k^2\left[\frac{D(k+q)-D(k)}{(k+q)^2-k^2}-\frac{\partial D(k)}{\partial k^2}\right],
\eea
which has the property of ensuring (by construction) that $\overline{T}(0)=0$, while, at the same time, 
coinciding with the original $T$ for all momenta $q$.

In addition, it is convenient to re-express $R(q)$ and $\overline{T}(q)$ 
in terms of the ghost dressing function. 
Using  Eq.~(\ref{ghdr}), 
after some elementary algebra, one obtains
\bea
R(q)&=&-\int_k\frac{F(k)}{k^2(k+q)^2}+\int_k\!\frac{F(k+q)-F(k)}{k^2[(k+q)^2-k^2]}, 
\nonumber \\
\overline{T}(q)&=& 
\int_k \left[\frac{F(k+q)-F(k)}{(k+q)^2-k^2}-\frac{\partial F(k)}{\partial k^2}\right];
\label{finalform}
\eea 
note that the angular integration of the first term in $R$ can be carried out 
analytically for any value of the space-time dimension $d$. 

Finally, note that up until this point we have been working in 
Minkowski space. To make the transition to Euclidean space, 
we must employ the usual rules. Specifically, 
we set $\int_k\!=\mathrm{i}\!\int_{k_\mathrm{\s E}}$ and $q^2_\mathrm{\s E} = -q^2$, 
and use that 
\be
\Delta_\mathrm{\s E}(q^2_\mathrm{\s E})=-\Delta(-q^2_\mathrm{\s E});
\qquad F_\mathrm{\s E}(q^2_\mathrm{\s E})=  F(-q^2_\mathrm{\s E});
\qquad G_\mathrm{\s E}(q^2_\mathrm{\s E})= G(-q^2_\mathrm{\s E}), 
\ee
suppressing the subscript ``E'' in what follows. 

\section{\label{num} Numerical evaluation of $\Pi_c(q^2)$ and $\Delta_r(q^2)$.}

We will now proceed to perform the numerical analysis. Using the available lattice data on the
ghost dressing function $F$,  we evaluate the terms $R$ and $\overline{T}$ given in 
Eq.~(\ref{finalform}),  
and combine them following the 
Eqs.~(\ref{PicTR}) and (\ref{rpic}) 
 to obtain the (renormalized) ghost contribution to the 
gluon self-energy $\Pi_c$ (of course, all relevant formulas must be properly ``euclideanized'').
Finally, we construct $\Delta_r$ using~(\ref{me}) and the
lattice results available for the gluon propagator $\Delta$. This exercise is carried out 
for three different cases: $d=4$ and $N=2,3$, as well as $d=3$ and $N=2$. 

\subsection{The case with $d=4$, $N=3$ }

\begin{figure}[!t]
\hspace{-1.5cm}
\begin{minipage}[b]{0.45\linewidth}
\centering
\includegraphics[scale=0.65]{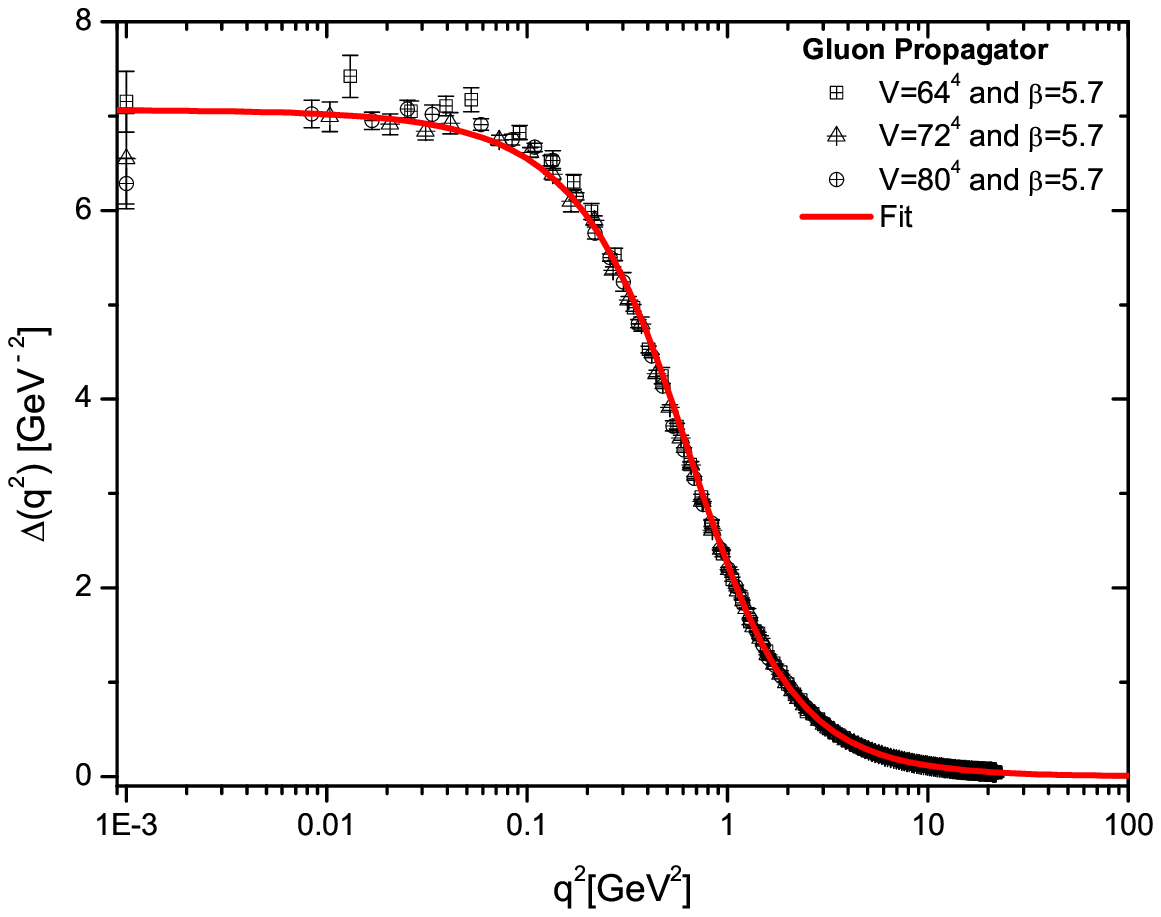}
\end{minipage}
\hspace{0.5cm}
\begin{minipage}[b]{0.50\linewidth}
\includegraphics[scale=0.65]{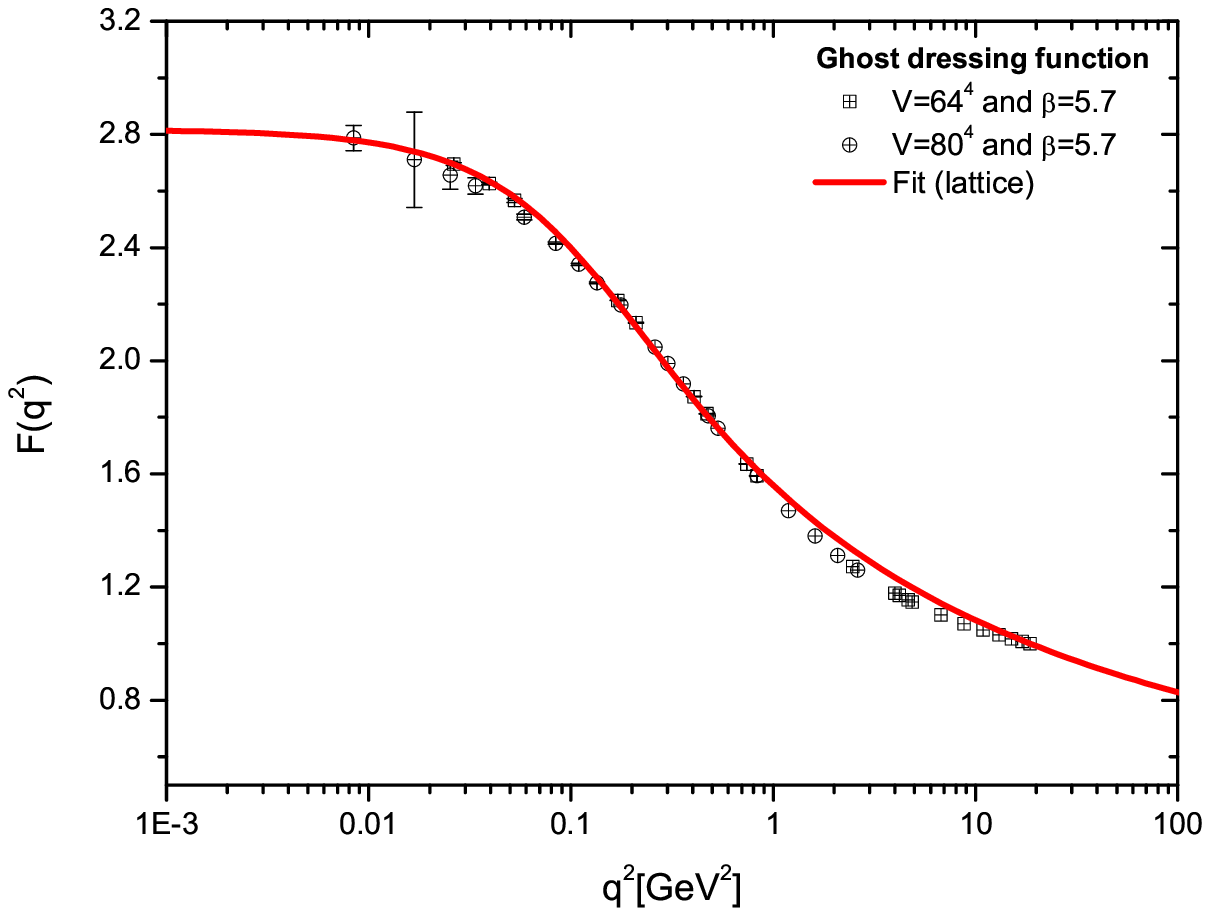}
\end{minipage}
\vspace{-0.5cm}
\caption{\label{ggh-4dSU3}{\it Left panel}: Lattice result for the $SU(3)$ gluon 
propagator, $\Delta(q)$, in $d=4$, renormalized  at \mbox{$\mu=4.3$ GeV}. The continuous 
line represents the fit given by Eq.~(\ref{gluon}).  {\it Right panel}: The $SU(3)$ ghost dressing
function, $F(q^2)$, renormalized at the same point, \mbox{$\mu=4.3$ GeV}; the solid line
corresponds to the fit given by Eq.~(\ref{ghdr-fit}).}
\end{figure}

In Fig.~\ref{ggh-4dSU3} we show the lattice results for the four-dimensional $SU(3)$ 
gluon propagator $\Delta(q^2)$ (left panel), and the  corresponding ghost dressing
function $F(q^2)$ (right panel), 
obtained from~\cite{Bogolubsky:2007ud}, and renormalized at 
$\mu=4.3$ GeV. 

As has been discussed in detail in the 
literature~\cite{Aguilar:2010cn,Aguilar:2011ux,Aguilar:2010gm}, both sets of data 
can be accurately fitted in terms of IR-finite quantities. 
More specifically, for the 
case of $\Delta(q^2)$, we have  proposed
a fit of the form~\cite{Aguilar:2010gm}
\be
\Delta^{-1}(q^2)= M^2(q^2) + q^2\left[1+ \frac{13C_{\rm A}g_1^2}{96\pi^2} 
\ln\left(\frac{q^2 +\rho_1\,M^2(q^2)}{\mu^2}\right)\right],
\label{gluon}
\ee  
where
\be
M^2(q^2) = \frac{m_0^4}{q^2 + \rho_2 m_0^2}.
\label{dmass}
\ee

Notice that in the above expression, the finiteness of $\Delta^{-1}(q^2)$ is assured  
by the presence of the function $M^2(q^2)$, which forces the value 
of \mbox{$\Delta^{-1}(0) = M^2(0) = m_0^2/\rho_2$}.
The continuous line on the left panel of Fig.~\ref{ggh-4dSU3} 
corresponds our best fit, which can be 
reproduced setting \mbox{$m_0 = 520$~MeV}, $g_1^2=5.68$, $\rho_1=8.55$  and  $\rho_2=1.91$.

The $SU(3)$ lattice data for $F(q^2)$, 
shown in the right panel of Fig.~\ref{ggh-4dSU3}, will be fitted by 
the following expression 
\be
F^{-1}(q^2)= 1+ \frac{9}{4}\frac{C_{\rm A}g_1^2}{48\pi^2}
\ln\left(\frac{q^2 +\rho_3  M^2(q^2)}{\mu^2}\right);\qquad M^2(q^2) = \frac{m_0^4}{q^2 + \rho_2 m_0^2},
\label{ghdr-fit}
\ee
with the parameters given by \mbox{$m_0 = 520$~MeV}, $g_2^2 = 8.65$, $\rho_2 = 0.68$ 
and $\rho_3 = 0.25$ . Notice that the $M(q^2)$ has the 
same power-law running as the one 
reported in Eq.~(\ref{dmass})~\cite{Lavelle:1991ve,Aguilar:2007ie,Oliveira:2010xc}.

It is interesting to notice that the aforementioned fits share 
the following important properties: {\it (i)} they 
connect smoothly the IR and UV regions by means of a unique expression; {\it (ii)} their 
finiteness is associated with the presence of the parameter $M$ in the argument of 
the perturbative (renormalization group) logarithm, which it is responsible
for taming the Landau pole and for doing 
the  logarithm saturates at a finite value~\cite{Aguilar:2010gm}; and {\it (iii)}  
for large values of $q^2$, Eqs.~(\ref{gluon}) and (\ref{ghdr-fit}) reproduce their 
respective one-loop expressions in the Landau gauge.

The only missing ingredient for the actual nonperturbative 
determination of $\Pi_c$, and therefore $\Delta_r$, is the value of \mbox{$\alpha_s=g^2/4\pi$}. 
Instead of choosing a single value for $\alpha_s$, we will establish a certain 
physically motivated range of values, which will furnish a more representative 
picture of the numerical impact of the ghost corrections on the gluon 
propagator. 
The lower value for $\alpha_s$ will be fixed simply by resorting to the  
the 4-loop (perturbative) calculations in the MOM scheme~\cite{Boucaud:2008gn}, 
and extracting the value of $\alpha_s$ that corresponds to the  subtraction point of 
\mbox{$\mu=4.3$ GeV}, used to renormalize the lattice data. 
The  value so obtained is $\alpha_s=0.2$. 

\begin{figure}[!t]
\includegraphics[scale=.65]{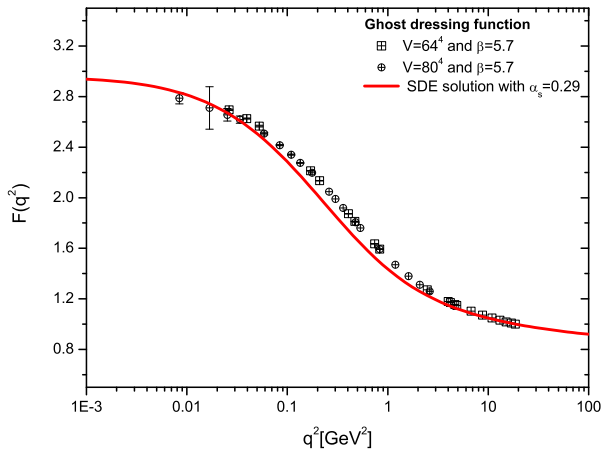}
\vspace{-0.5cm}
\caption{\label{alpha_s-4dSU3} The solution of the SDE~(\ref{ghsde}) 
that best matches the ghost dressing function data is obtained for $\alpha_s=0.29$.}
\end{figure}

In order to establish a reasonable upper bound, 
in a consistent way, we resort to the methodology employed 
in~\cite{Aguilar:2009nf}, which makes use of the standard SDE for the 
ghost dressing function, given by (Euclidean space)
\be
F^{-1}(q^2) = 1 +g^2 C_{\rm {A}} \int_k \frac1{(k+q)^2}
\left[1-\frac{(k\cdot q)^2}{k^2q^2}\right] \Delta (k)  F(k+q),
\label{ghsde}
\ee
derived in the Landau gauge, and under the assumption 
that the full ghost-gluon vertex is 
approximated by its tree-level value~\cite{Aguilar:2009nf,Cucchieri:2004sq}. 
In this integral equation one substitutes for $\Delta (k)$ the fit given in  Eq.~(\ref{gluon}),
and solves it numerically for the unknown function $F(q^2)$; evidently, for each value of 
$\alpha_s$ we obtain a different solution for $F(q^2)$.
The correct value of $\alpha_s$ is then determined as the one for which the 
corresponding (renormalized) solution best matches the 
lattice results~(see Fig.~\ref{alpha_s-4dSU3}); for \mbox{$\mu=4.3$ GeV} we 
obtain $\alpha_s=0.29$, showing that the perturbative MOM value ($\alpha_s=0.2$) is 30\% lower. 

\begin{figure}[!t]
\hspace{-1.5cm}
\begin{minipage}[b]{0.45\linewidth}
\centering
\includegraphics[scale=0.65]{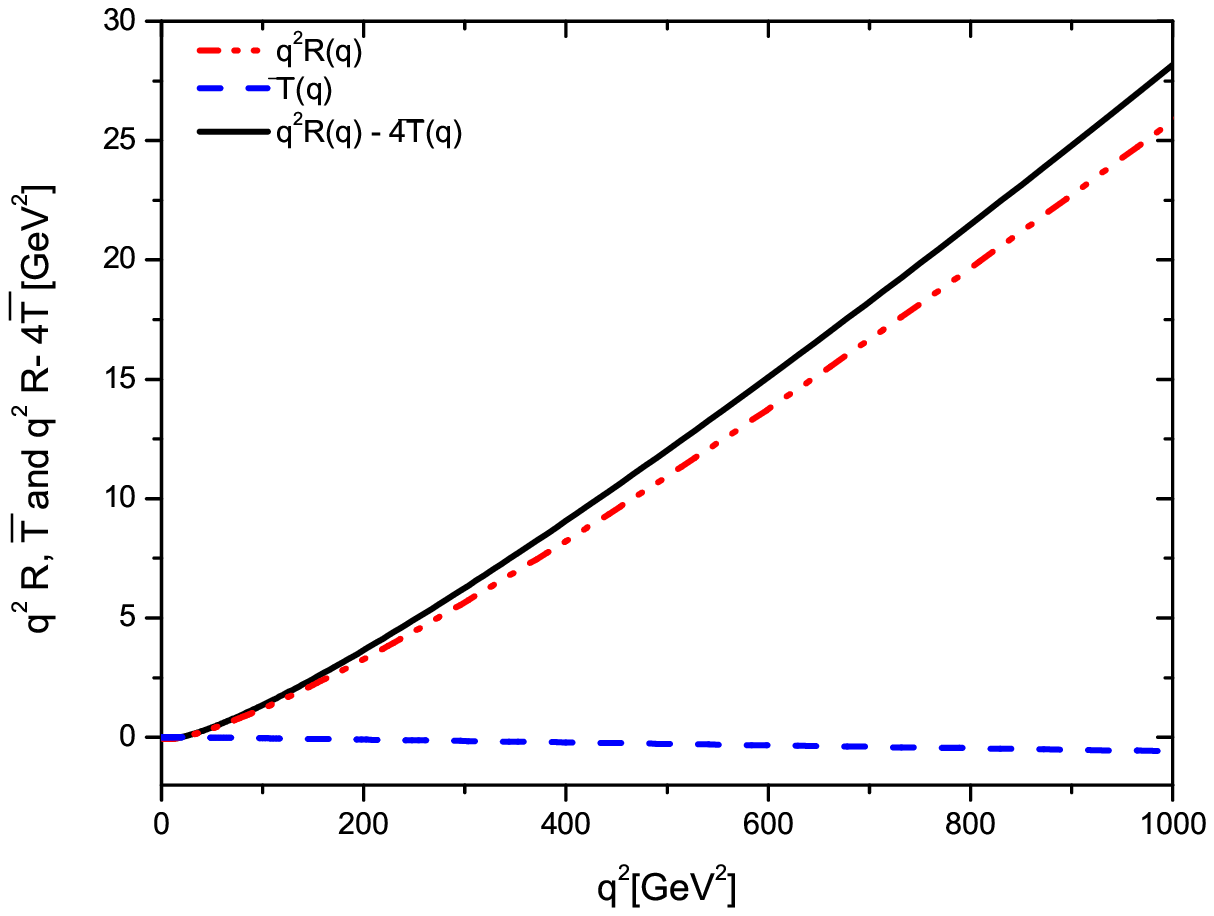}
\end{minipage}
\hspace{0.5cm}
\begin{minipage}[b]{0.50\linewidth}
\includegraphics[scale=0.65]{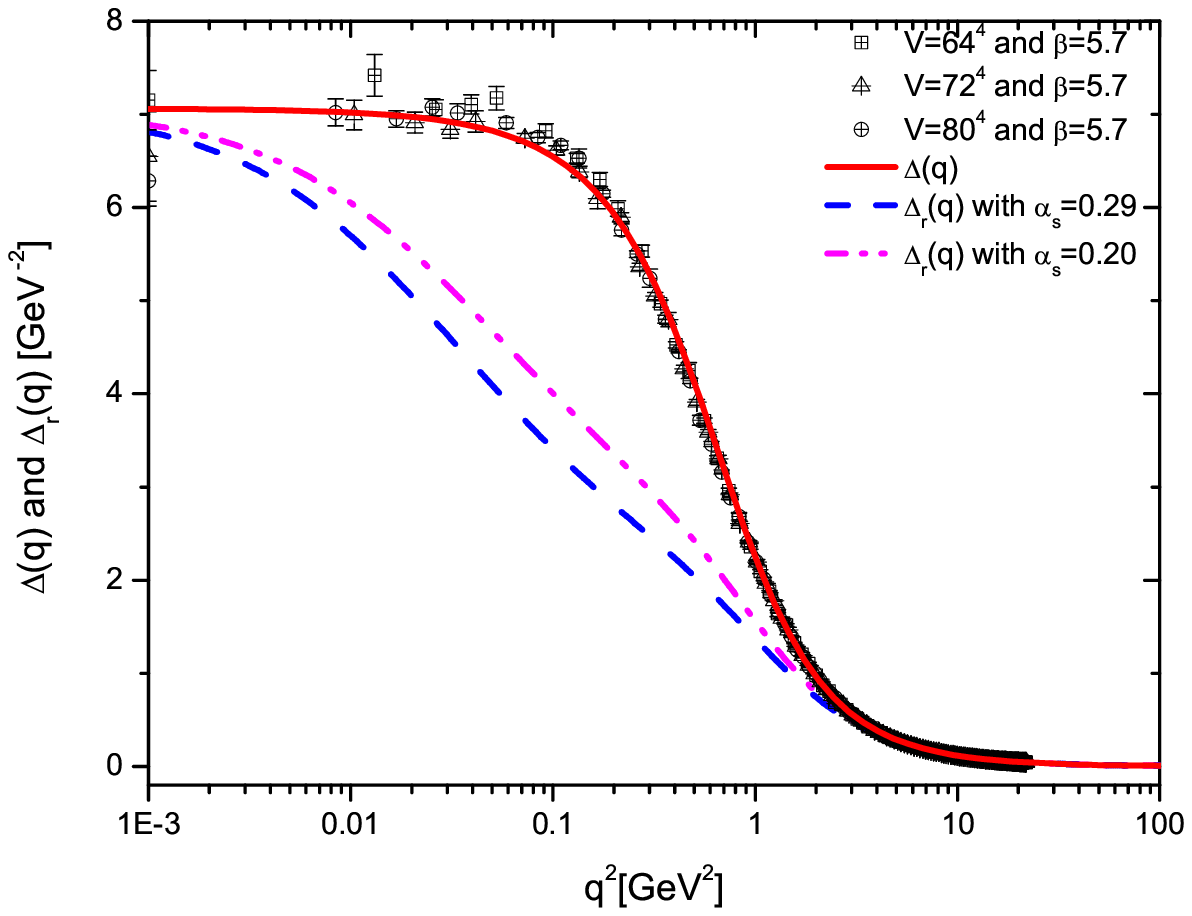}
\end{minipage}
\vspace{-0.5cm}
\caption{\label{RTandGluon-4dSU3}{\it Left panel}: Numerical evaluation 
of the ghost contribution $\Pi_c(q)$ to the gluon propagator using as input
 our best fit for the $d=4$, $N=3$ ghost dressing lattice data. {\it Right panel}: 
The removal of the one-loop dressed ghost contribution from the (lattice) 
gluon propagator  results in a diminished ``swelling'' in the momentum region below 1 GeV$^2$.} 
\label{GdressandChi-4dSU3}
\end{figure}

The results obtained for the renormalized $R$ and $\overline{T}$,  
after substituting into the corresponding formulas our best fit
for $F$, given by Eq.~(\ref{ghdr-fit}), are shown on the left panel of Fig.~\ref{RTandGluon-4dSU3}, together 
with the combination  $q^2 R -4 \overline{T}$, which appears on the rhs of 
Eq.~(\ref{PicTR}). It is clear that the contribution of the term $4 \overline{T}$ 
is rather negligible; in a way this is to be expected, given that this term 
vanishes identically in perturbation theory (for all values of $q$), and 
vanishes nonperturbatively at the origin [{\it viz.} Eqs.(\ref{RandTpert}) and~(\ref{T0}), respectively].   

Next, we use these results to construct $\Pi_c$, given in Eq.~(\ref{rpic}), and 
finally $\Delta_r$, expressed by Eq.~(\ref{rme}) (Fig.~\ref{RTandGluon-4dSU3} right panel), using both 
values of $\alpha_s$, namely \mbox{$\alpha_s=0.29$} (SDE, red dotted line) and 
\mbox{$\alpha_s=0.20$} (4-loop MOM, blue dashed-dotted line). 

We then see that the net effect of removing the ghost contribution is to suppress 
significantly  
the support of the gluon propagator in the region below \mbox{1 GeV$^2$}. 
Higher values of $\alpha_s$  increase the impact of the ghost contributions, 
but only slightly,  as can be seen on the right panel of Fig.~\ref{RTandGluon-4dSU3}. 
As we will see in the next section, this ``deflating'' of the gluon propagator 
in the intermediate region of momenta, 
produced by the removal of the ghost contributions,  
has far-reaching consequences on the generation of a dynamical gluon mass.

\subsection{The case with $d=4$, $N=2$}

\begin{figure}[!t]
\hspace{-1.5cm}
\begin{minipage}[b]{0.45\linewidth}
\centering
\includegraphics[scale=0.65]{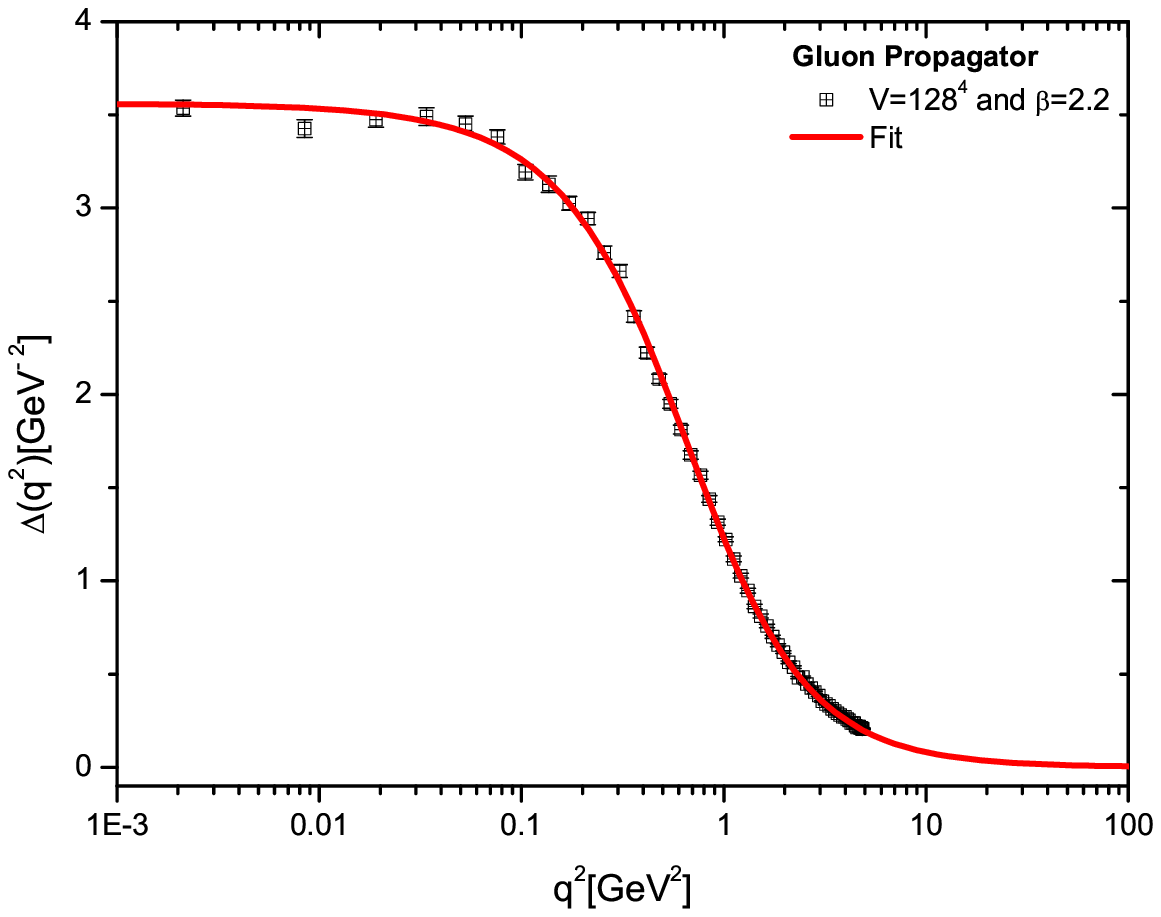}
\end{minipage}
\hspace{0.5cm}
\begin{minipage}[b]{0.50\linewidth}
\includegraphics[scale=0.65]{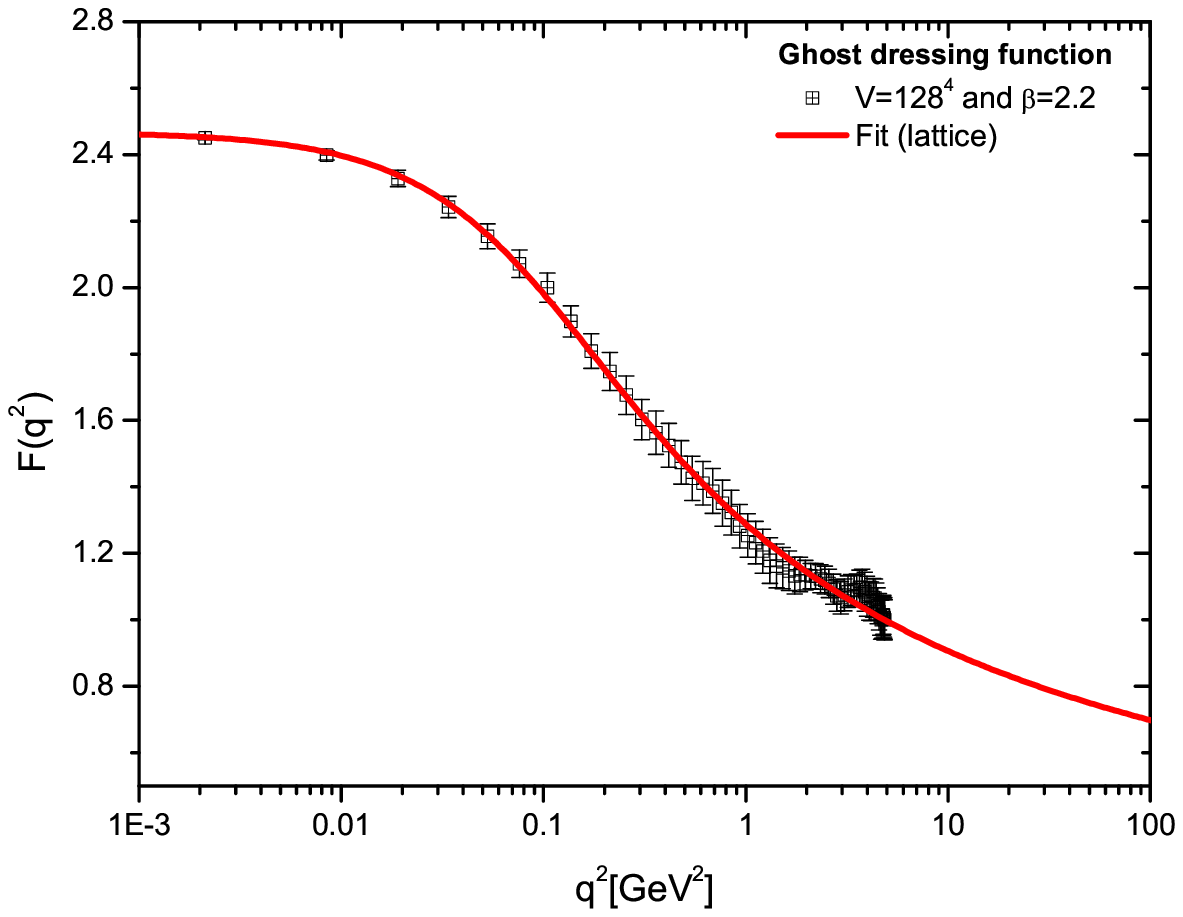}
\end{minipage}
\vspace{-0.5cm}
\caption{\label{ggh-4dSU2}{\it Left panel}: Lattice results for the $SU(2)$ gluon 
propagator in $d=4$, renormalized  at $\mu=2.2$ GeV. The continuous 
line represents our best fit to the data
obtained from Eq.~(\ref{gluon}).  {\it Right panel}: The $SU(2)$ ghost dressing
function $F(q^2)$, renormalized at the same point, $\mu=2.2$ GeV; the solid line
corresponds to the best fit given by Eq.~(\ref{ghdr-fit}).} 
\end{figure}

It turns out that, changing the gauge group  to $SU(2)$ does not significantly alter
the characteristic qualitative behavior found in the $SU(3)$ case. Specifically, in Fig.~\ref{ggh-4dSU2} we
show the gluon propagator (left panel), and the ghost dressing 
function (right panel),  obtained from~\cite{Cucchieri:2010xr} and renormalized at $\mu=2.2$ GeV. 

As in the $SU(3)$ case, the gluon and ghost data can be accurately fitted by the expressions~(\ref{gluon}) and~(\ref{ghdr-fit}), where now $C_A=2$ and the fitting parameters are \mbox{$m_0= 865$\,\mbox{MeV}}, \mbox{$g_1^2=10.80$}, \mbox{$\rho_1=1.96$} and, \mbox{$\rho_2=2.68$} (gluon) and  \mbox{$g_2^2 = 15.03$},
\mbox{$m_0 = 523\,$ MeV} \mbox{$\rho_3=0.215$} and \mbox{$\rho_4=0.781$} (ghost).

\begin{figure}[!t]
\includegraphics[scale=.65]{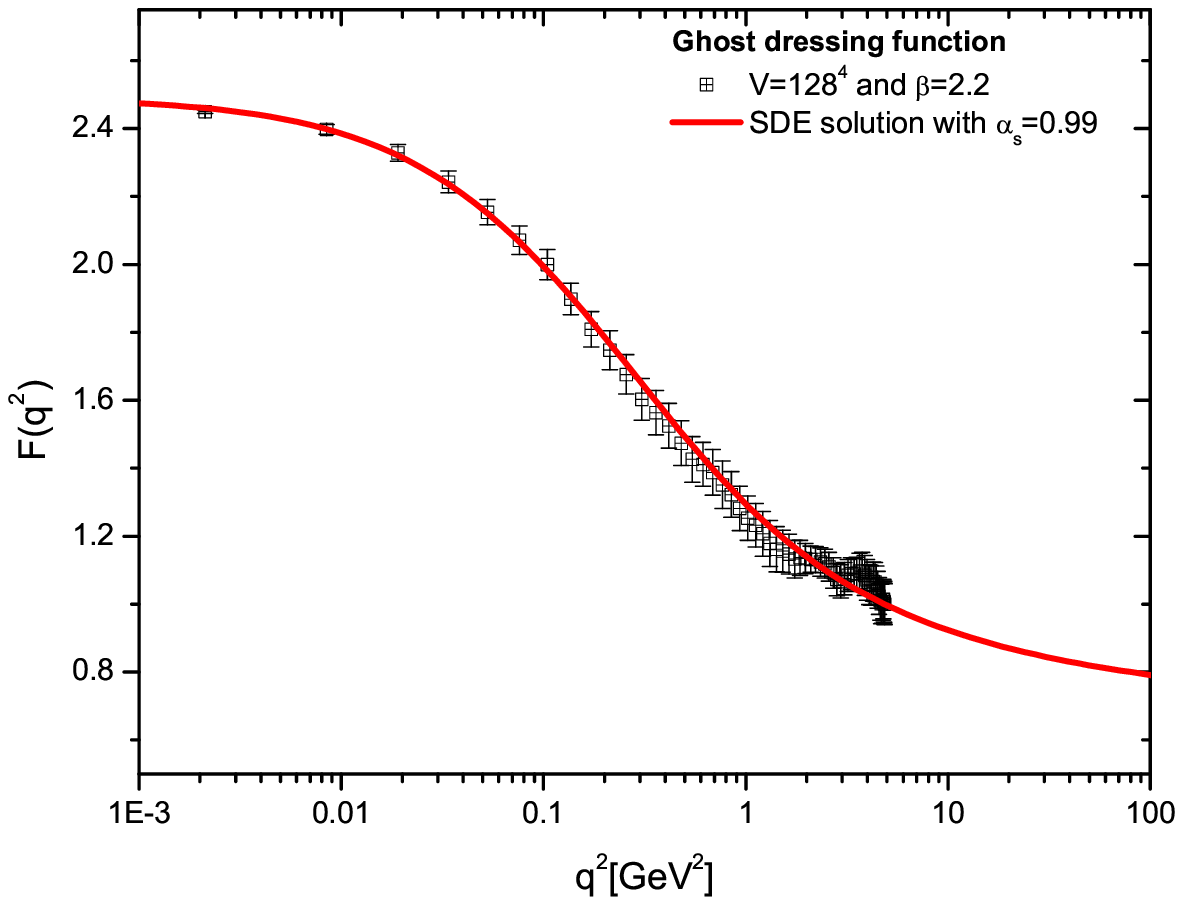}
\vspace{-0.5cm}
\caption{\label{alpha_s-4dSU2}  The solution of the SDE~(\ref{ghsde}) that 
best matches the ghost dressing function data is obtained for \mbox{$\alpha_s=0.99$}.}
\end{figure}

The coupling $\alpha_s$ can be also fixed using 
the same procedure described in the previous subsection
(Fig.~\ref{alpha_s-4dSU2}); the value obtained from the SD solution
that best matches the lattice data is in this case is \mbox{$\alpha_s=0.99$}.

\begin{figure}[!t]
\hspace{-1.5cm}
\begin{minipage}[b]{0.45\linewidth}
\centering
\includegraphics[scale=0.65]{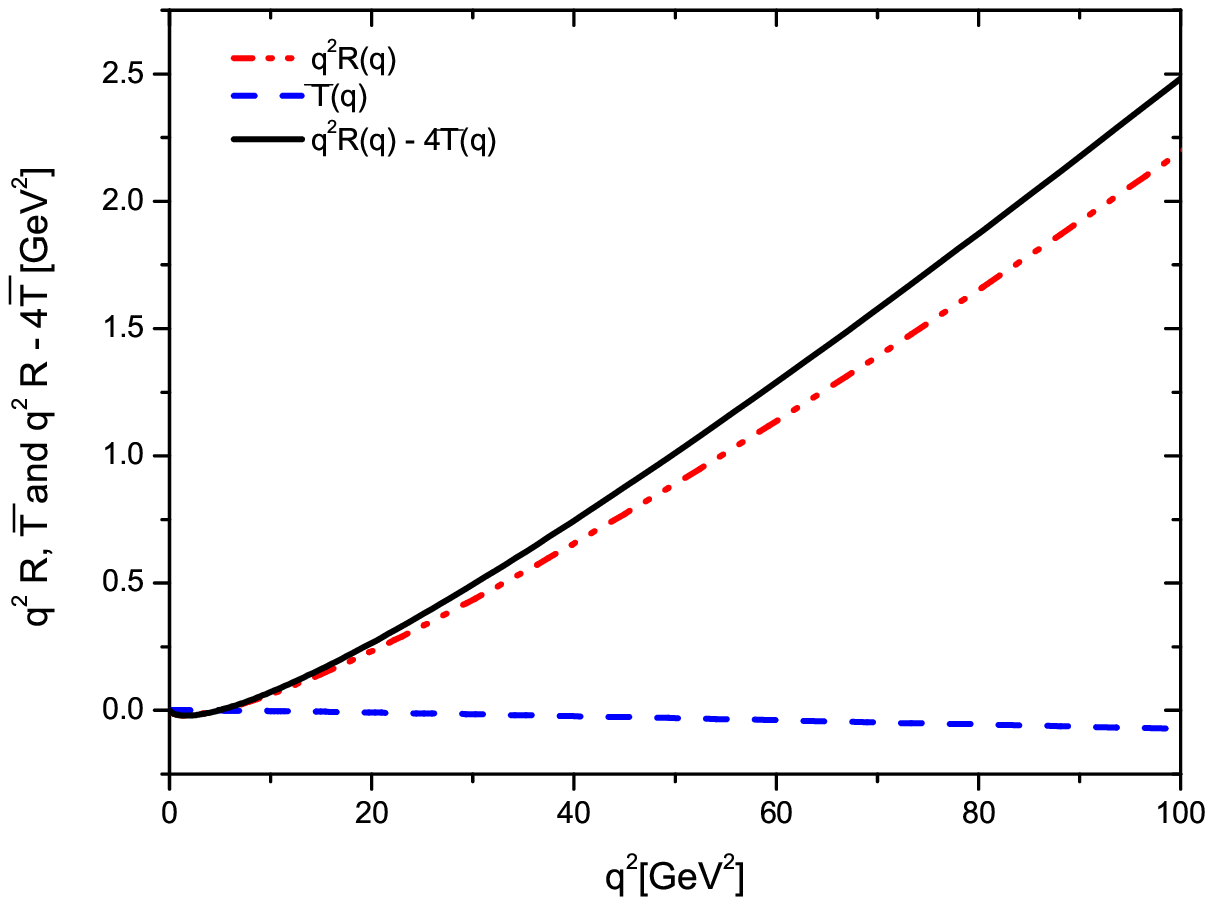}
\end{minipage}
\hspace{0.5cm}
\begin{minipage}[b]{0.5\linewidth}
\includegraphics[scale=0.65]{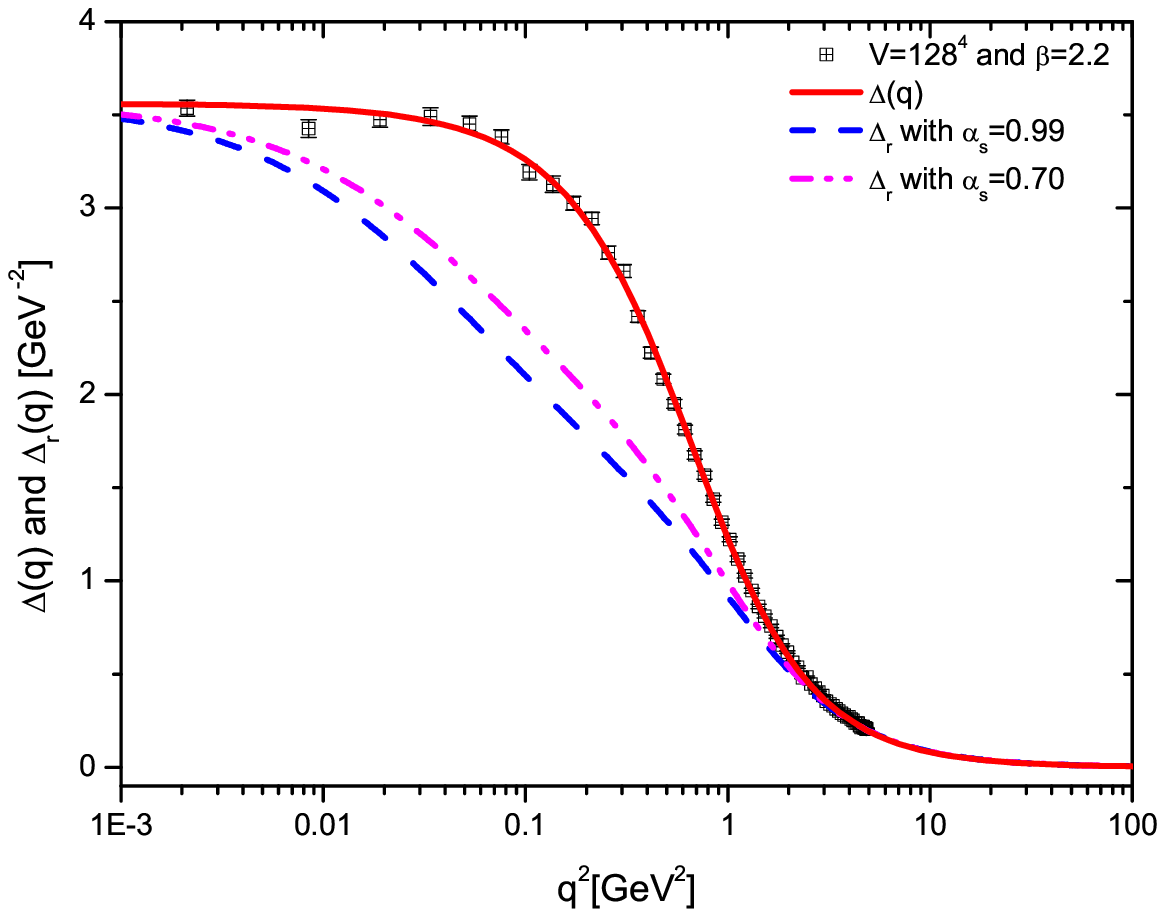}
\end{minipage}
\vspace{-0.5cm}
\caption{\label{RTandGluon-4dSU2}{\it Left panel}: Numerical evaluation of 
the ghost contribution $\Pi_c$ to the gluon propagator using as input our 
best fit~(\ref{ghdr-fit}) for the $d=4$, $N=2$ ghost dressing lattice data.
{\it Right panel}: The removal of the one-loop dressed ghost contribution from 
the (lattice) gluon propagator  causes, as in the $SU(3)$ case, a 
considerable suppression in the momentum region below 1 GeV$^2$.}
\end{figure}

On the left panel of Fig.~\ref{RTandGluon-4dSU2}, we show the resulting 
curves for $R$ and $\overline{T}$ obtained through our best fit for $F$ given by Eq.~(\ref{ghdr-fit}).
Then, using Eqs.~(\ref{PicTR}) and (\ref{rme}) we combine the previous results to 
get the $SU(2)$ ghost self-energy $\Pi_c$ and, finally, $\Delta_r$ 
(right panel of the same figure). We use again two values for $\alpha_s$ namely 
the one obtained through the solution of the ghost SDE (\mbox{$\alpha_s=0.99$}) and 
a 30\% lower one (\mbox{$\alpha_s=0.70$}). Evidently, the $SU(2)$ results do not differ qualitatively 
from those of the  $SU(3)$ case: a lower value for $\alpha_s$ suppresses the ghost contribution 
to the gluon propagator, and the removal of the ghost gives rise to a  
lower curve in the region below 1 GeV$^2$.

\subsection{The case with $d=3$, $N=2$}

\begin{figure}[!t]
\hspace{-1.5cm}
\begin{minipage}[b]{0.45\linewidth}
\centering
\includegraphics[scale=0.65]{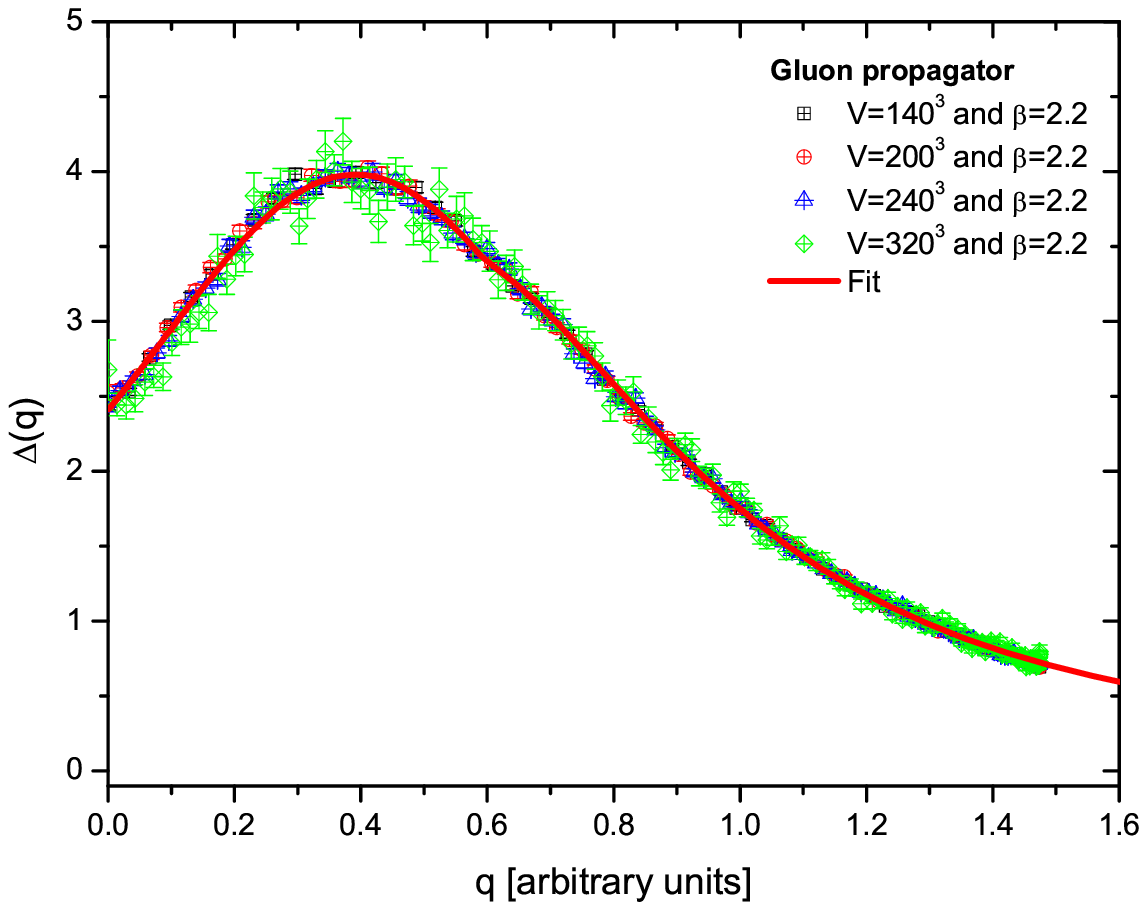}
\end{minipage}
\hspace{0.5cm}
\begin{minipage}[b]{0.50\linewidth}
\includegraphics[scale=0.65]{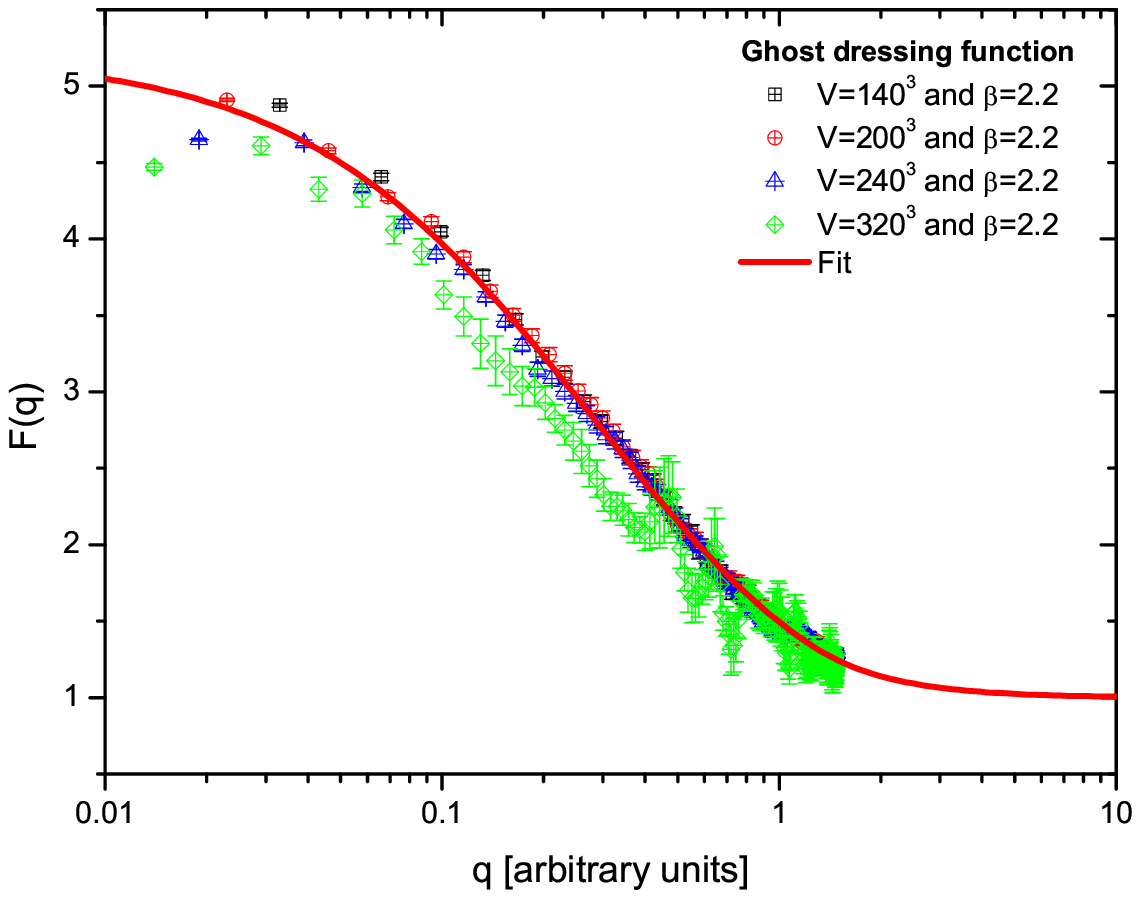}
\end{minipage}
\vspace{-0.5cm}
\caption{{\it Left panel}: Lattice results for the $SU(2)$ gluon 
propagator in $d=3$. The continuous 
line represents our best fit to the data
obtained from Eq.~(\ref{fit3d}).  {\it Right panel}: Lattice data for the $SU$(2) ghost 
dressing function $F(q)$ in 3 dimensions; the solid line 
corresponds to the best fit given by Eq.~(\ref{ghdr3d-fit}).} 
\label{GdressandChi-3d}
\end{figure}

Let us start, as in the previous cases,  by showing in Fig.~\ref{GdressandChi-3d} 
the lattice results~\cite{Cucchieri:2003di,Cucchieri:2010xr}  for the three-dimensional gluon 
propagator $\Delta(q)$ (left panel) and the  ghost dressing function $F(q)$ (right panel).
Notice that, in Fig.~\ref{GdressandChi-3d}, the lattice data for $\Delta(q)$  presented 
in Ref.~\cite{Cucchieri:2003di,Cucchieri:2010xr} were appropriately
rescaled, following the procedure explained in detail in~\cite{Aguilar:2010zx}, 
 to match correctly the perturbative tail. Both $\Delta(q)$ and $F(q)$
saturate in the deep  IR region, and can therefore be fitted by means of IR finite expressions. 

In the case of the gluon propagator, an accurate fit is giving by 
\be
\Delta(q) = A\exp\left[-(q-q_0)^2/w\right] + \frac{1}{a+bq + cq^2} \,, 
\label{fit3d}
\ee
where the fitting parameters 
are $A=0.49$, $q_0=0.11$, $w=0.37$, $a=0.43$, $b=-0.85$, and $c=1.143$. 
For the ghost dressing function, 
we use the following piecewise interpolator
\bea
F(q) &=& \frac1{a+bq+cq^2}, \,\,\, \mbox{for} \quad q^2 \leq 3 
\nonumber \\ 
&=&  1 + \frac{d}{eq + q^2},  \,\,\, \mbox{for} \quad q^2 > 3
\label{ghdr3d-fit}
\eea
with fitting coefficients $a=0.19$, $b=0.61$, $c= -0.14$, $d=0.63$ and $e=0.26$ 
obtained by requiring the function to be continuous at $q^2=3$.

\begin{figure}[!t]
\includegraphics[scale=.65]{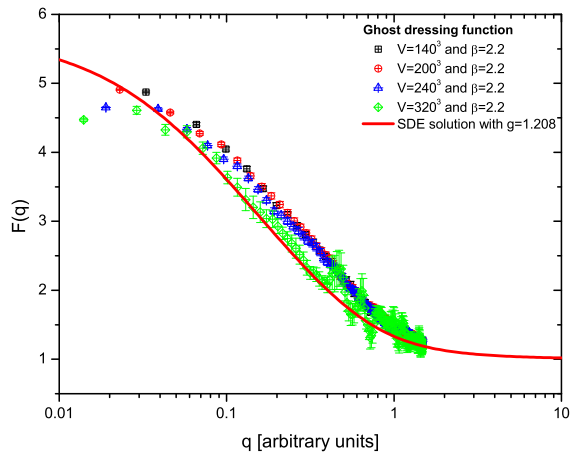}
\vspace{-0.5cm}
\caption{\label{alpha_s-3}  The solution of the SDE~(\ref{ghsde}) that 
best matches the ghost dressing function data in $d=3$ is obtained for \mbox{$g=1.208$}.}
\end{figure}

\begin{figure}[!t]
\hspace{-1.5cm}
\begin{minipage}[b]{0.45\linewidth}
\centering
\includegraphics[scale=0.65]{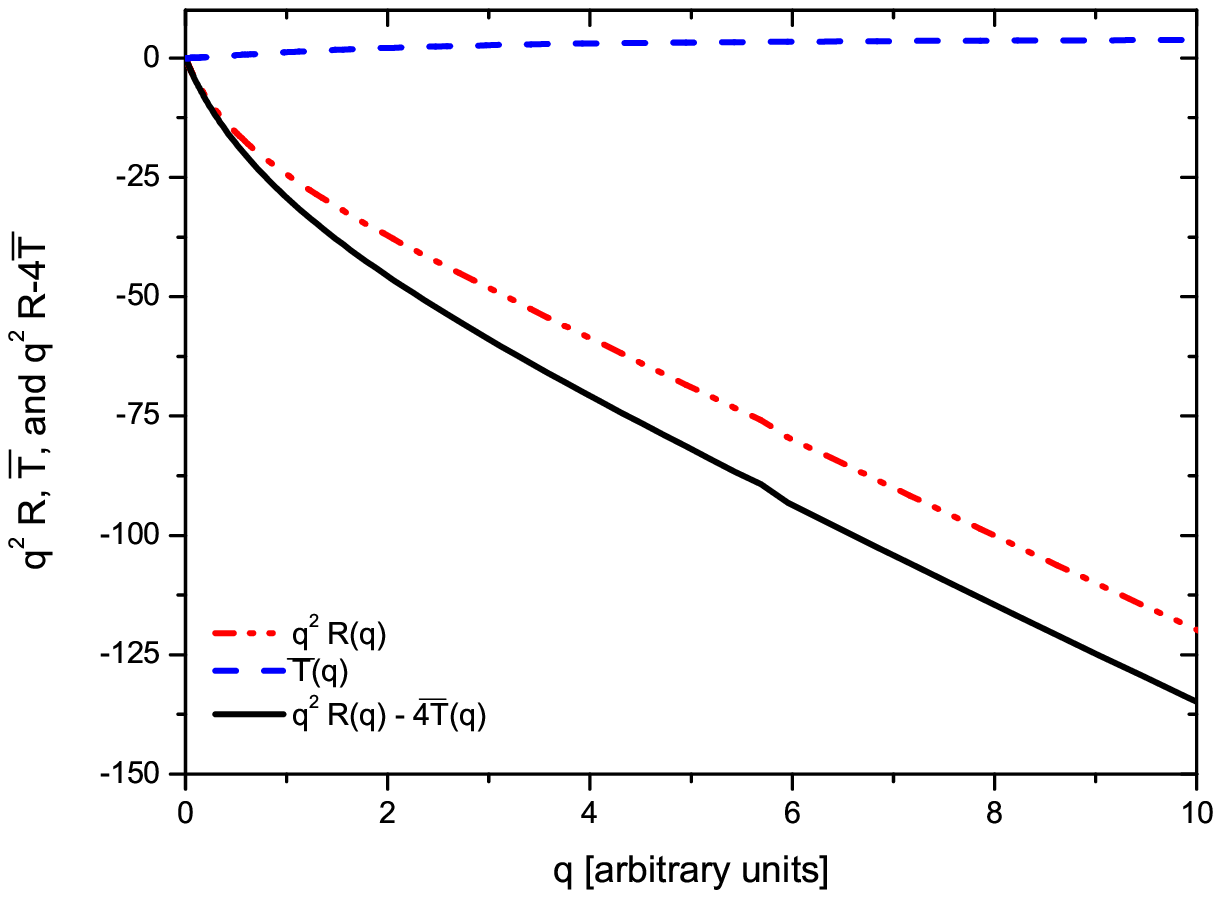}
\end{minipage}
\hspace{0.5cm}
\begin{minipage}[b]{0.50\linewidth}
\includegraphics[scale=0.65]{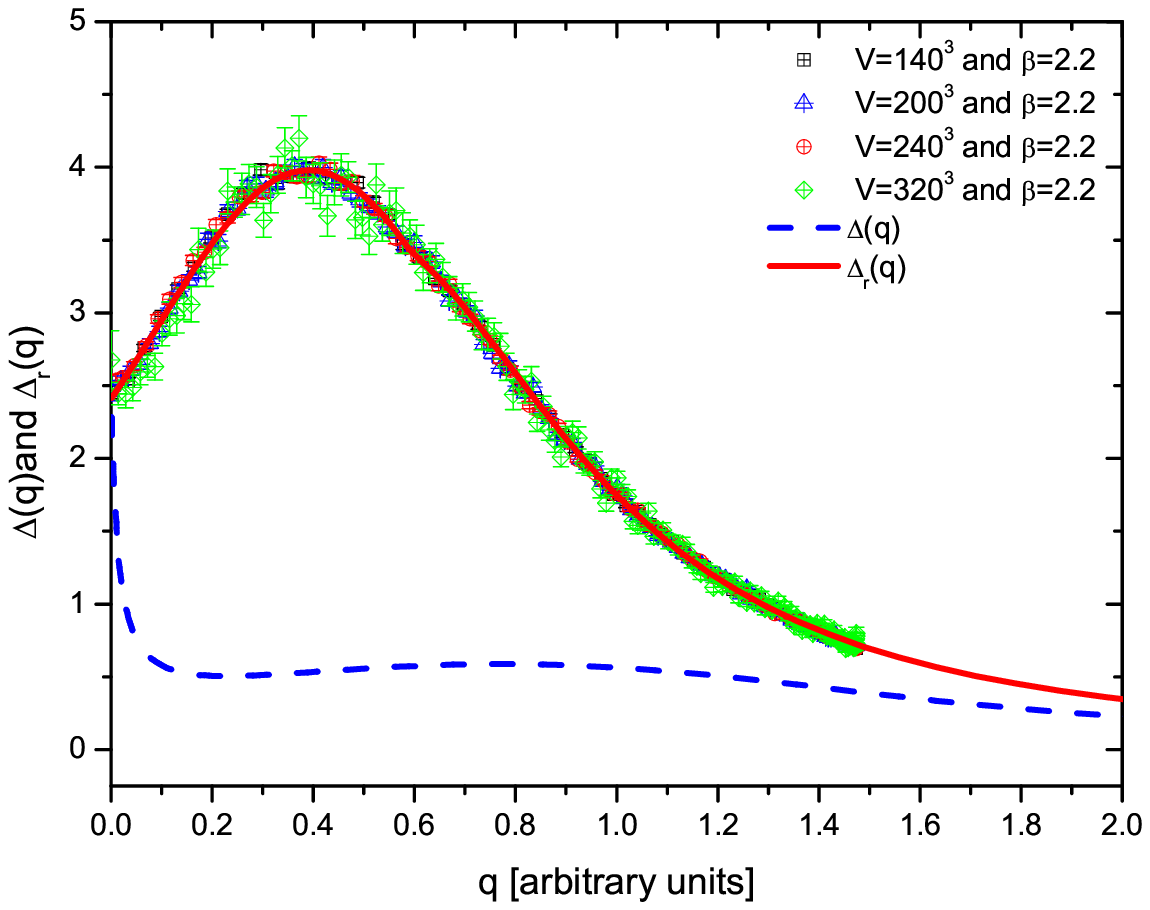}
\end{minipage}
\vspace{-0.5cm}
\caption{\label{RTandGluon-3d}{\it Left panel}: Numerical evaluation 
of the ghost contribution $\Pi_c$ to the gluon propagator using as 
input our best fit for the $d=3$, $N=2$ ghost dressing lattice 
data. {\it Right panel}: The result of removing the one-loop dressed 
ghost contribution from the gluon propagator in $d=3$. The effect is 
much more dramatic than in the $d=4$ case, since all the structure 
is determined by the ghost contribution, while $\Delta_r$ has the sole 
(but crucial!) role of rendering the propagator finite at $q=0$.} 
\end{figure}

The contribution of $R$ and $\overline{T}$ of Eq.~(\ref{finalform}) can  
be then evaluated using the above fit, and the results of this calculation are shown in 
the left panel of Fig.~\ref{RTandGluon-3d}. Since $d=3$ Yang-Mills is a 
super-renormalizable theory, 
all aforementioned quantities are directly UV finite, and do not need to undergo renormalization. 

The next step is to determine the value of the coupling constant $g$ (which, in $d=3$, 
has  dimensions of $m^{1/2}$) entering in
the formulas for  $\Pi_c$ and $\Delta_r$, given by Eqs.~(\ref{PicTR}) and (\ref{me}), respectively. 
The procedure followed is the same as before, {\it i.e.} we will employ the three-dimensional ghost SDE, 
solve it for various values of $g$, and choose the one that best reproduces the lattice data for $F$.
The most favorable case is shown in Fig.~\ref{alpha_s-3}, where 
the solution for $F(q)$ obtained from the SDE with $g=1.208$ 
[in the same arbitrary mass units used in the plots of Fig.~\ref{GdressandChi-3d}] (red line) 
is compared with the lattice results for the same quantity. 

Next, substituting the results presented on the left panel of  Fig.~\ref{RTandGluon-3d}
into Eqs.~(\ref{PicTR}) and (\ref{me}), and using  \mbox{$g=1.208$}, we compute $\Pi_c$ and $\Delta_r$.
On the left panel of Fig.~\ref{RTandGluon-3d}, 
we compare the residual propagator $\Delta_r$ (blue dashed line) with the full propagator $\Delta(q)$. Clearly, the 
effect in the tridimensional case 
is even more pronounced: the 
ghost contribution completely dominates over the rest, 
determining to a large extent the overall shape and structure of the propagator.

\section{\label{DMG} No gluon mass without ghost loops}

In the previous section we have studied how the subtraction 
of the ghost contributions affects the profile of the  gluon
propagator. However, as we will now show, the effects goes way  
beyond a simple change in the overall propagator shape, modifying its salient 
qualitative characteristics, and in particular the generation of a dynamical gluon mass.

To establish this, we start from the dynamical equation
describing the effective gluon mass, recently 
derived in~\cite{Aguilar:2011ux}; it reads (Euclidean space)
\be
m^2(q^2)=\frac{2g^2C_A}{1+G(q^2)}\int_k\!\left[k^2-\frac{(k\cdot q)^2}{q^2}\right]
\frac{m^2(k+q)-m^2(k)}{(k+q)^2-k^2}\Delta(k)\Delta(k+q).
\label{me-final}
\ee
Taking the $q\to0$ limit, one then gets
\bea
m^2(0)&=&\frac{2g^2C_A}{1+G(0)}\frac{d-1}d\int_k\!k^2[m^2(k)]'\Delta^2(k)\nonumber\\
&=&-\frac{4g^2C_A}{1+G(0)}\frac{d-1}d\int_k\!m^2(k)\Delta(k)\left[k^2\Delta(k)\right]',
\label{0cond}
\eea
where in the last step we have used integration by parts. Introducing 
spherical coordinates (setting $y=k^2$) and the $d$-dimensional integral
measure [notice that in~(\ref{me-final}) there is no dependence
on the $d-2$ polar angles $\varphi_i$]
\be
\int_k\ =\ \frac1{(2\pi)^d}\frac{\pi^{\frac{d-1}2}}{\Gamma\left
(\frac{d-1}2\right)}\int_0^\pi\!\diff\theta\sin^{d-2}\theta\int_0^\infty\!\diff y\,y^{\frac d2-1},
\label{d-measure}
\ee
Eq.~(\ref{0cond}) finally becomes
\be
m^2(0)=-\frac{d-1}{d(4\pi)^{\frac d2}\Gamma
\left(\frac d2\right)}\frac{4g^2C_A}{1+G(0)}\int_0^\infty\!\diff y\,m^2(y){\cal K}_{d;N}(y),
\label{fnal}
\ee
with the kernel ${\cal K}_{d;N}$ given by
\be
{\cal K}_{d;N}(y)=y^{\frac d2-1}\Delta(y)[y\Delta(y)]'. 
\label{ker}
\ee
The dependence of ${\cal K}_{d;N}$ 
on $N$ (the number of colors) is implicit in the form of $\Delta(y)$ that must be employed in each case, 
{\it i.e.}, $\Delta(y)=\Delta_N(y)$. The same is true for $G(0)$ in (\ref{fnal}), and, of course, $C_A=N$.  

\begin{figure}[!t]
\includegraphics[scale=1.15]{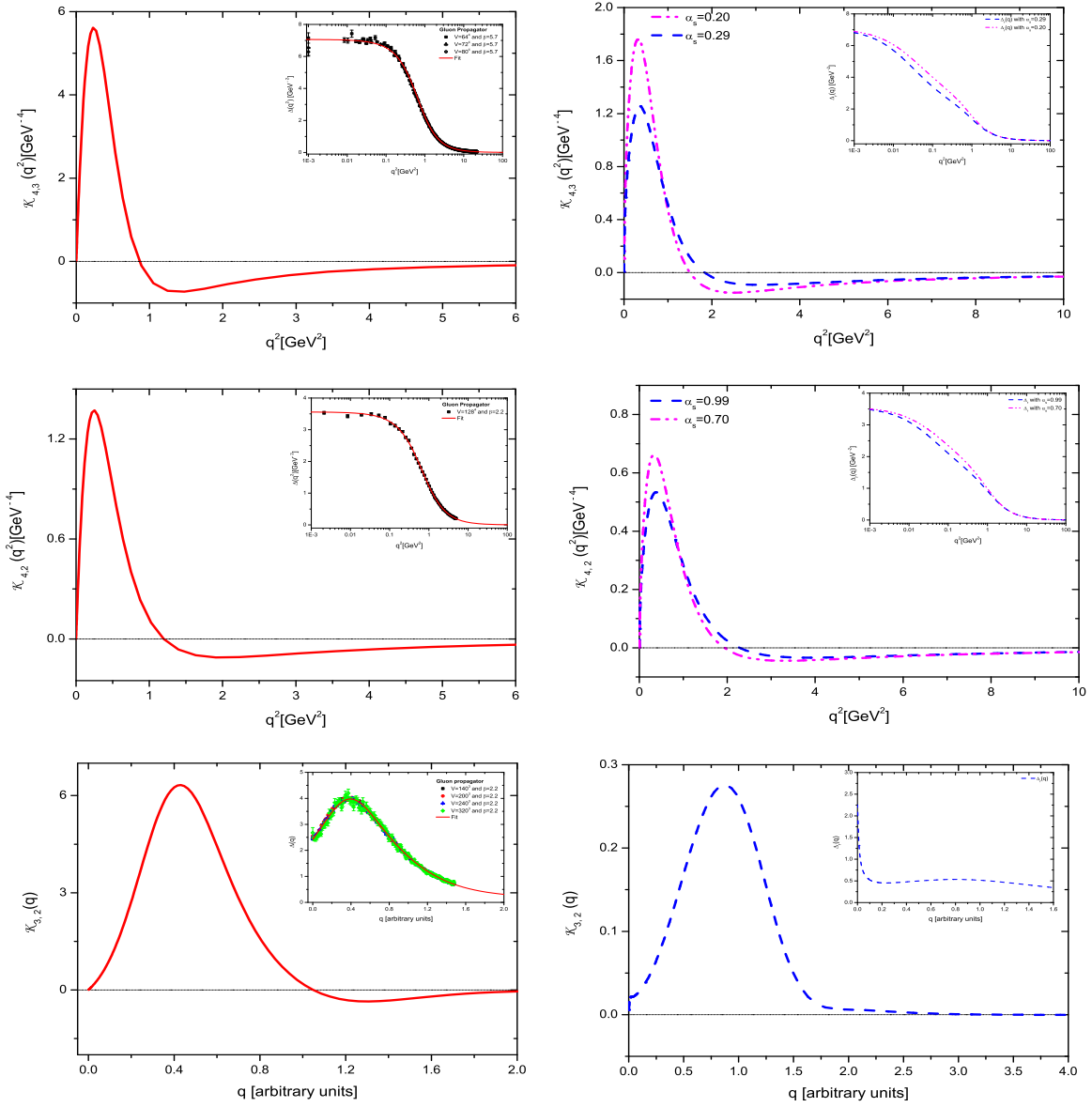}
\caption{\label{Kernel-all}The kernel ${\cal K}_{d;N}$ of Eq.~(\ref{ker}) 
constructed out of the lattice propagator $\Delta$ (left panels) and
the ghost-less propagator $\Delta_r$ (right panels) for the $d=4$ $N=3$ 
(top row), $N=2$ (middle row) and  $d=3$ $N=2$ (bottom row) cases. The insets
show in each case the shape of the propagator used to evaluate the kernels.}
\end{figure}

Since the constant multiplying the integral is positive, the 
negative sign in front of Eq.~(\ref{fnal}) tells us that the 
required physical constraint $m^2(0)>0$ can be fulfilled if and only 
if the integral kernel ${\cal K}_{d;N}$ (constructed solely out
of the gluon propagator) displays a sufficiently deep and extended negative
region at intermediate momenta~\cite{Aguilar:2011ux}.

In the left panels of Fig.~\ref{Kernel-all} we plot the 
kernels ${\cal K}_{d;N}$ obtained from the lattice data for the
cases  $d=4$, $N=3$  (top row) and $N=2$ (middle row), as
well as  $d=3$ $N=2$ (bottom row), considered in the previous section; they 
all posses the characteristic negative region that allows, at least
in principle, the existence of solutions of Eq.~(\ref{me-final}), furnishing 
a positive value for the condition~(\ref{0cond}). We emphasize that,  
for the $d=4$ and $N=3,2$ cases such a solution has been explicitly found
and studied in~\cite{Aguilar:2011ux}. Notice the striking resemblance 
between the kernels obtained for the different cases.

On the other hand, the situation changes substantially once 
the ghost loop is removed, in which case the kernels ${\cal K}_{d;N}$ must be 
constructed from $\Delta_r$ (right panels of the same figure). For $d=4$ one 
observes a shift towards higher $q$s of the zero crossing, and a correspondingly
suppressed negative region; even though this is not sufficient to
exclude {\it per se} the existence of a physical solution to the mass
equation~(\ref{me-final}), a thorough study of the approximate
equation derived in~\cite{Aguilar:2011ux} reveals that no physical solution 
may be found. The $d=3$ situation is even more obvious: the highly suppressed negative region present 
in this case cannot support solutions of~(\ref{0cond}) with $m^2(0)>0$, thus
leaving as the only possibility the trivial $m^2=0$ solution.

The main conclusion one can draw, therefore, is that the ghosts play 
a fundamental role in the mechanism of 
dynamical gluon mass generation, since the failure to
properly include them results in the inability of the theory to generate
dynamically a mass for the gluon. This, in turn, implies that what is displayed in the right
panels of Figs.~\ref{RTandGluon-4dSU3}, \ref{RTandGluon-4dSU2},~\ref{RTandGluon-3d}, 
and \ref{Kernel-all} are not the gluon propagators one would actually obtain, 
assuming that one were actually able to perform this ``experiment'', {\it e.g.}, remove the ghosts on the 
lattice. Indeed,  according to our results, if ghosts were not included, $m^2=0$, and thus $\Delta$ would 
not saturate in the IR at all!

To understand what happens, let us concentrate on the $d=4$, $N=3$ case and
imagine a simplified setting, where one can switch off adiabatically 
the ghosts, neglecting all other effects this operation would entail 
(we will come back to this point at the end of this section). This 
could be achieved by multiplying the  self-energy $\Pi_c$, appearing in Eq.~(\ref{gSDE}), by
a parameter $\gamma\in[0,1]$, such that when $\gamma=1$ the full $\Delta$ of the
right panel of  Fig.~\ref{RTandGluon-4dSU3} is reproduced. 
Now, by slowly decreasing $\gamma$ (for fixed $\alpha$) one would give 
rise to a set of intermediate $\Delta_{r,\gamma}$ profiles, showing 
progressively less ``swelling'' in the $q^2<1$ [GeV$^2$] region, and ideally
one would get $\Delta_{r,0}\equiv\Delta_r$. However  before that will
happen, there will exist a critical value $\gamma_c$ for which the 
kernel ${\cal K}_{4;3}$ constructed from $\Delta_{r,\gamma_c}$ will fail 
to provide the required negative region that would ensure the positivity 
of $m^2(0)$, as calculated from the condition~(\ref{fnal}). At that 
point the theory will undergo a drastic change, showing a gluon
propagator that does not saturate in the IR. Even though we cannot 
actually predict what such a propagator might behave like in the deep IR, it is likely that 
the typical singularity associated with the  (perturbative) Landau  pole 
(tamed by the presence of the mass) may reappear.

Obviously in this analysis we are neglecting  any type 
of back-reaction due to the changes in the gluon propagator: to
be sure, any modification to the latter quantity would affect 
not only the ghost -- since  the gluon propagator appears in 
fact in the ghost SDE, see Eq.~(\ref{ghsde}) -- but also the
gluon mass, and therefore the IR saturation value -- through
Eqs.~(\ref{me-final}) and (\ref{0cond}). While such 
effects might be numerically appreciable  
(changing, {\it e.g.}, the critical 
value $\gamma_c$), we expect the qualitative description 
given above to persist.

\section{\label{concl}Conclusions}

In this article we have presented 
a detailed study of the impact of the ghost sector  
on the overall form of the gluon propagator in a pure Yang-Mills theory,   
for different space-time dimensions ($d=3,\ 4$) and $SU(N)$ gauge groups ($N=2,3$).

The key ingredients for performing this analysis have been basically two. To begin 
with, the PT-BFM framework allowed us to subtract out {\it gauge-invariantly} the 
``one-loop dressed'' ghost diagrams from the SDE describing the full  
 gluon propagator.
Second, we have been able to express these ghost  
contributions as a simple integral involving the ghost dressing function 
only. This was achieved by employing a judicious Ansatz for the ghost-gluon 
vertex, obtained by solving the corresponding Ward identity, and  
by resorting to the ``seagull-identity'', in order to enforce certain crucial properties.
The nonperturbative evaluation of the resulting expressions have been 
carried out numerically, using available lattice data as input for the ghost dressing function.
Our results reveal that the (``one-loop dressed'') ghost diagrams furnish 
a sizable contribution to the gluon propagator in $d=4$, and the dominant one 
in $d=3$.  

The suppression of the gluon propagator induced by the removal of the ghost-loops 
has far-reaching consequences on the mechanism that endows gluons with a 
dynamical mass, associated with the  observed IR-finiteness of 
the gluon propagator and the ghost-dressing function.
Specifically, using a recently 
derived integral equation controlling the dynamics of the (momentum-dependent) gluon mass,
we have demonstrated that when the  reduced gluon propagators are used as inputs, 
the corresponding kernels are  modified in such a way that no 
physical solutions may be found, thus failing to generate a mass gap for 
the pure Yang-Mills theory. Instead, as has been shown in~\cite{Aguilar:2011ux}, the 
use of the full gluon propagator in the same equation generates a physically acceptable 
gluon mass.   

Once the results of the present work are combined with those  
of~\cite{Aguilar:2010cn} for the chiral symmetry breaking, 
a compelling picture of QCD emerges, where the generation of a dynamical 
mass for quarks {\it and} gluons requires the synergistic participation 
of all fields (physical and unphysical) of the theory.

\acknowledgments 

The research of J.~P. is supported by the European FEDER and  Spanish MICINN under 
grant FPA2008-02878. The work of  A.C.A  is supported by the Brazilian
Funding Agency CNPq under the grant 305850/2009-1 and project 474826/2010-4 .


\begin{thebibliography}{99}



\bibitem{Cucchieri:2007md}
A.~Cucchieri and T.~Mendes,
PoS {\bf LAT2007}, 297 (2007).

\bibitem{Cucchieri:2007rg}
A.~Cucchieri and T.~Mendes,
Phys.\ Rev.\ Lett.\  {\bf 100}, 241601 (2008).

\bibitem{Cucchieri:2009zt}
A.~Cucchieri and T.~Mendes,
Phys.\ Rev.\  D {\bf 81}, 016005 (2010).

\bibitem{Cucchieri:2011ga}
 A.~Cucchieri and T.~Mendes,
PoS {\bf LATTICE2010}, 280 (2010).

\bibitem{Cucchieri:2011um}
  A.~Cucchieri, T.~Mendes,
  AIP Conf.\ Proc.\  {\bf 1343}, 185-187 (2011).

\bibitem{Cucchieri:2003di}
  A.~Cucchieri, T.~Mendes and A.~R.~Taurines,
  Phys.\ Rev.\  D {\bf 67}, 091502 (2003).

\bibitem{Cucchieri:2010xr}
 A.~Cucchieri and T.~Mendes,
  PoS {\bf QCD-TNT09}, 026 (2009)


\bibitem{Bowman:2007du}
P.~O.~Bowman {\it et al.},
Phys.\ Rev.\  D {\bf 76}, 094505 (2007).


\bibitem{Bogolubsky:2007ud}
 I.~L.~Bogolubsky, E.~M.~Ilgenfritz, M.~Muller-Preussker and A.~Sternbeck,
PoS {LATTICE}, 290 (2007).

\bibitem{Bogolubsky:2009dc}
 I.~L.~Bogolubsky, E.~M.~Ilgenfritz, M.~Muller-Preussker and A.~Sternbeck,
Phys.\ Lett.\  B {\bf 676}, 69 (2009).

\bibitem{Oliveira:2008uf}
  O.~Oliveira, P.~J.~Silva,
  Phys.\ Rev.\  {\bf D79}, 031501 (2009).

\bibitem{Oliveira:2009eh}
O.~Oliveira and P.~J.~Silva,
PoS {\bf LAT2009}, 226 (2009).


\bibitem{Aguilar:2008xm}
  A.~C.~Aguilar, D.~Binosi and J.~Papavassiliou,
  Phys.\ Rev.\  D {\bf 78}, 025010 (2008).

\bibitem{Aguilar:2010zx}
  A.~C.~Aguilar, D.~Binosi and J.~Papavassiliou,
  Phys.\ Rev.\  D {\bf 81}, 125025 (2010).

\bibitem{Binosi:2009qm}  
D.~Binosi and J.~Papavassiliou,
Phys.\ Rept.\  {\bf 479}, 1-152 (2009).

  
\bibitem{RodriguezQuintero:2011vw}
  J.~Rodriguez-Quintero,
  Phys.\ Rev.\  {\bf D83}, 097501 (2011).

\bibitem{RodriguezQuintero:2010wy}
  J.~Rodriguez-Quintero,
  JHEP {\bf 1101}, 105 (2011).
   
 
\bibitem{Boucaud:2010gr}
  Ph.~Boucaud, M.~E.~Gomez, J.~P.~Leroy, A.~Le Yaouanc, J.~Micheli, O.~Pene, J.~Rodriguez-Quintero,
  Phys.\ Rev.\  {\bf D82}, 054007 (2010).


\bibitem{Boucaud:2008gn}
  Ph.~Boucaud, F.~De Soto, J.~P.~Leroy, A.~Le Yaouanc, J.~Micheli, O.~Pene and J.~Rodriguez-Quintero,
  Phys.\ Rev.\  D {\bf 79}, 014508 (2009).

\bibitem{Boucaud:2008ji}
Ph.~Boucaud, J.~P.~Leroy, A.~L.~Yaouanc, J.~Micheli, O.~Pene and J.~Rodriguez-Quintero,
JHEP {\bf 0806} (2008) 012.

\bibitem{Braun:2007bx}
J.~Braun, H.~Gies and J.~M.~Pawlowski,
Phys.\ Lett.\  B {\bf 684}, 262 (2010).

\bibitem{Szczepaniak:2010fe}
A.~P.~Szczepaniak and H.~H.~Matevosyan,
Phys.\ Rev.\  D {\bf 81}, 094007 (2010).


\bibitem{Dudal:2008sp}
D.~Dudal, J.~A.~Gracey, S.~P.~Sorella, N.~Vandersickel and H.~Verschelde,
Phys.\ Rev.\  D {\bf 78}, 065047 (2008).


\bibitem{Dudal:2010tf}
  D.~Dudal, O.~Oliveira, N.~Vandersickel,
  Phys.\ Rev.\  {\bf D81}, 074505 (2010).

\bibitem{Dudal:2011gd}
  D.~Dudal, S.~P.~Sorella, N.~Vandersickel,
  [arXiv:1105.3371 [hep-th]].

\bibitem{Kondo:2011ab}
  K.~-I.~Kondo,
  [arXiv:1103.3829 [hep-th]].


\bibitem{Jackiw:1973tr}
  R.~Jackiw, K.~Johnson,
  Phys.\ Rev.\  {\bf D8}, 2386-2398 (1973).
  
\bibitem{Cornwall:1973ts}
  J.~M.~Cornwall, R.~E.~Norton,
  Phys.\ Rev.\  {\bf D8}, 3338-3346 (1973).
  

\bibitem{Eichten:1974et}
  E.~Eichten, F.~Feinberg,
  Phys.\ Rev.\  {\bf D10}, 3254-3279 (1974).
  

\bibitem{Cornwall:1981zr}
J.~M.~Cornwall,
Phys.\ Rev.\ D {\bf 26}, 1453 (1982).


\bibitem{Bernard:1981pg}
  C.~W.~Bernard,
  Phys.\ Lett.\  {\bf B108}, 431 (1982); 
  Nucl.\ Phys.\  {\bf B219}, 341 (1983).
    

\bibitem{Alkofer:2000wg}
 R.~Alkofer, L.~von Smekal,
 Phys.\ Rept.\  {\bf 353}, 281 (2001).

\bibitem{Fischer:2006ub}
 C.~S.~Fischer,
 J.\ Phys.\ G {\bf G32}, R253-R291 (2006).


\bibitem{Fischer:2003rp}
  C.~S.~Fischer, R.~Alkofer,
  Phys.\ Rev.\  {\bf D67}, 094020 (2003).


\bibitem{Aguilar:2010cn}
A.~C.~Aguilar and J.~Papavassiliou,
Phys.\ Rev.\ D {\bf 83}, 014013 (2011).


\bibitem{Roberts:1994dr}
  C.~D.~Roberts and A.~G.~Williams,
  Prog.\ Part.\ Nucl.\ Phys.\  {\bf 33}, 477 (1994).


\bibitem{Cornwall:1989gv}
J.~M.~Cornwall and J.~Papavassiliou,
Phys.\ Rev.\  D {\bf 40}, 3474 (1989).

\bibitem{Binosi:2002ft}
D.~Binosi and J.~Papavassiliou,
Phys.\ Rev.\  D {\bf 66}(R), 111901 (2002).

\bibitem{Binosi:2003rr}
D.~Binosi and J.~Papavassiliou,
J.\ Phys.\ G {\bf 30}, 203 (2004).


\bibitem{Abbott:1980hw}
See, e.g.,  L.~F.~Abbott,
Nucl.\ Phys.\  B {\bf 185}, 189 (1981), and references therein.

\bibitem{Aguilar:2006gr}
A.~C.~Aguilar and J.~Papavassiliou,
JHEP {\bf 0612}, 012 (2006).

\bibitem{Binosi:2007pi}
D.~Binosi and J.~Papavassiliou,
Phys.\ Rev.\  D {\bf 77}(R), 061702 (2008).

\bibitem{Binosi:2008qk}
D.~Binosi and J.~Papavassiliou,
JHEP {\bf 0811}, 063 (2008).

\bibitem{Aguilar:2009ke}
  A.~C.~Aguilar and J.~Papavassiliou,
  Phys.\ Rev.\  D {\bf 81}, 034003 (2010).

\bibitem{Aguilar:2011ux}
  A.~C.~Aguilar, D.~Binosi and J.~Papavassiliou,
  arXiv:1107.3968 [hep-ph].

\bibitem{Grassi:2004yq}
  P.~A.~Grassi, T.~Hurth and A.~Quadri,
  Phys.\ Rev.\  D {\bf 70}, 105014 (2004).


\bibitem{Aguilar:2009pp}
  A.~C.~Aguilar, D.~Binosi and J.~Papavassiliou,
  JHEP {\bf 0911}, 066 (2009).




\bibitem{Salam:1963sa}
  A.~Salam,
  Phys.\ Rev.\  {\bf 130}, 1287 (1963);
  A.~Salam and R.~Delbourgo,
  Phys.\ Rev.\  {\bf 135}, B1398 (1964);
  R.~Delbourgo and P.~C.~West,
  J.\ Phys.\ A  {\bf 10}, 1049 (1977);
  R.~Delbourgo and P.~C.~West,
  Phys.\ Lett.\  B {\bf 72}, 96 (1977).

\bibitem{Ball:1980ay}
  J.~S.~Ball, T.~-W.~Chiu,
  Phys.\ Rev.\  {\bf D22}, 2542 (1980).

\bibitem{Kizilersu:2009kg}
  A.~Kizilersu and M.~R.~Pennington,
  Phys.\ Rev.\  D {\bf 79}, 125020 (2009);
  A.~Bashir, A.~Kizilersu and M.~R.~Pennington,
  Phys.\ Rev.\  D {\bf 57}, 1242 (1998).



\bibitem{Curtis:1990zs}
  D.~C.~Curtis and M.~R.~Pennington,
  Phys.\ Rev.\  D {\bf 42}, 4165 (1990).







\bibitem{Aguilar:2010gm}
  A.~C.~Aguilar, D.~Binosi, J.~Papavassiliou,
  JHEP {\bf 1007}, 002 (2010).


\bibitem{Lavelle:1991ve}
  M.~Lavelle,
  Phys.\ Rev.\  {\bf D44}, 26-28 (1991).

\bibitem{Aguilar:2007ie}
  A.~C.~Aguilar, J.~Papavassiliou,
  Eur.\ Phys.\ J.\  {\bf A35}, 189-205 (2008).

\bibitem{Oliveira:2010xc}
  O.~Oliveira, P.~Bicudo,
  J.\ Phys.\ G {\bf G38}, 045003 (2011).

\bibitem{Aguilar:2009nf}
  A.~C.~Aguilar, D.~Binosi, J.~Papavassiliou and J.~Rodriguez-Quintero,
  Phys.\ Rev.\  D {\bf 80}, 085018 (2009)
  
  
\bibitem{Cucchieri:2004sq}
  A.~Cucchieri, T.~Mendes and A.~Mihara,
  JHEP {\bf 0412}, 012 (2004).
  



\end{thebibliography}
\end{document}